\documentclass[longauth,traditabstract]{aa}

\usepackage[nonamebreak]{natbib}
\usepackage[stable]{footmisc}
\usepackage{amsmath}
\usepackage{txfonts}
\usepackage{placeins}
\usepackage{natbib}
\bibpunct{(}{)}{;}{a}{}{,} %
\usepackage{graphicx}
\usepackage{epstopdf}
\usepackage{ifthen}
\usepackage[breaklinks, colorlinks, citecolor=blue]{hyperref}
\usepackage{overpic}
\usepackage{fixltx2e}
\usepackage{rotating}
\usepackage{mathtools}
\usepackage{kbordermatrix}
\usepackage{ulem}
\usepackage[table,usenames,dvipsnames]{xcolor}

\newcommand{\StokesI}{I}                    
\newcommand{\StokesQ}{Q}                    
\newcommand{\StokesU}{U}                    

\newcommand{\polangsky}{\gamma}             

\def\setsymbol#1#2{\expandafter\def\csname #1\endcsname{#2}}
\def\getsymbol#1{\csname #1\endcsname}

\def\Planck{\textit{Planck}}

\newbox\tablebox    \newdimen\tablewidth
\def\leaderfil{\leaders\hbox to 5pt{\hss.\hss}\hfil}
\def\tablenote#1 #2\par{\begingroup \parindent=0.8em
    \abovedisplayshortskip=0pt\belowdisplayshortskip=0pt
    \noindent
    $$\hss\vbox{\hsize\tablewidth \hangindent=\parindent \hangafter=1 \noindent
    \hbox to \parindent{$^#1$\hss}\strut#2\strut\par}\hss$$
    \endgroup}

\def\L2{\ifmmode L_2\else $L_2$\fi}

\def\DeltaT{\ifmmode \Delta T\else $\Delta T$\fi}
\def\deltat{\ifmmode \Delta t\else $\Delta t$\fi}
\def\fknee{\ifmmode f_{\rm knee}\else $f_{\rm knee}$\fi}
\def\Fmax{\ifmmode F_{\rm max}\else $F_{\rm max}$\fi}
\def\solar{\ifmmode{\rm M}_{\mathord\odot}\else${\rm M}_{\mathord\odot}$\fi}
\def\Msolar{\ifmmode{\rm M}_{\mathord\odot}\else${\rm M}_{\mathord\odot}$\fi}
\def\Lsolar{\ifmmode{\rm L}_{\mathord\odot}\else${\rm L}_{\mathord\odot}$\fi}
\def\inv{\ifmmode^{-1}\else$^{-1}$\fi}
\def\mo{\ifmmode^{-1}\else$^{-1}$\fi}
\def\sup#1{\ifmmode ^{\rm #1}\else $^{\rm #1}$\fi}
\def\expo#1{\ifmmode \times 10^{#1}\else $\times 10^{#1}$\fi}
\def\,{\thinspace}
\def\lsim{\mathrel{\raise .4ex\hbox{\rlap{$<$}\lower 1.2ex\hbox{$\sim$}}}}
\def\gsim{\mathrel{\raise .4ex\hbox{\rlap{$>$}\lower 1.2ex\hbox{$\sim$}}}}

\def\simprop{\mathrel{\raise .4ex\hbox{\rlap{$\propto$}\lower 1.2ex\hbox{$\sim$}}}}
\def\deg{\ifmmode^\circ\else$^\circ$\fi}
\def\pdeg{\ifmmode $\setbox0=\hbox{$^{\circ}$}\rlap{\hskip.11\wd0 .}$^{\circ}
          \else \setbox0=\hbox{$^{\circ}$}\rlap{\hskip.11\wd0 .}$^{\circ}$\fi}
\def\arcs{\ifmmode {^{\scriptstyle\prime\prime}}
          \else $^{\scriptstyle\prime\prime}$\fi}
\def\arcm{\ifmmode {^{\scriptstyle\prime}}
          \else $^{\scriptstyle\prime}$\fi}
\newdimen\sa  \newdimen\sb
\def\parcs{\sa=.07em \sb=.03em
     \ifmmode \hbox{\rlap{.}}^{\scriptstyle\prime\kern -\sb\prime}\hbox{\kern -\sa}
     \else \rlap{.}$^{\scriptstyle\prime\kern -\sb\prime}$\kern -\sa\fi}
\def\parcm{\sa=.08em \sb=.03em
     \ifmmode \hbox{\rlap{.}\kern\sa}^{\scriptstyle\prime}\hbox{\kern-\sb}
     \else \rlap{.}\kern\sa$^{\scriptstyle\prime}$\kern-\sb\fi}
\def\ra[#1 #2 #3.#4]{#1\sup{h}#2\sup{m}#3\sup{s}\llap.#4}
\def\dec[#1 #2 #3.#4]{#1\deg#2\arcm#3\arcs\llap.#4}
\def\deco[#1 #2 #3]{#1\deg#2\arcm#3\arcs}
\def\rra[#1 #2]{#1\sup{h}#2\sup{m}}

\def\dots{\relax\ifmmode \ldots\else $\ldots$\fi}
\def\WHzsr{\ifmmode $W\,Hz\mo\,sr\mo$\else W\,Hz\mo\,sr\mo\fi}
\def\mHz{\ifmmode $\,mHz$\else \,mHz\fi}
\def\GHz{\ifmmode $\,GHz$\else \,GHz\fi}
\def\mKs{\ifmmode $\,mK\,s$^{1/2}\else \,mK\,s$^{1/2}$\fi}
\def\muKs{\ifmmode \,\mu$K\,s$^{1/2}\else \,$\mu$K\,s$^{1/2}$\fi}
\def\muKRJs{\ifmmode \,\mu$K$_{\rm RJ}$\,s$^{1/2}\else \,$\mu$K$_{\rm RJ}$\,s$^{1/2}$\fi}
\def\muKHz{\ifmmode \,\mu$K\,Hz$^{-1/2}\else \,$\mu$K\,Hz$^{-1/2}$\fi}
\def\MJysr{\ifmmode \,$MJy\,sr\mo$\else \,MJy\,sr\mo\fi}
\def\MJysrmK{\ifmmode \,$MJy\,sr\mo$\,mK$_{\rm CMB}\mo$\else \,MJy\,sr\mo\,mK$_{\rm CMB}\mo$\fi}
\def\microns{\ifmmode \,\mu$m$\else \,$\mu$m\fi}

\def\muK{\ifmmode \,\mu$K$\else \,$\mu$\hbox{K}\fi}
\def\microK{\ifmmode \,\mu$K$\else \,$\mu$\hbox{K}\fi}
\def\muW{\ifmmode \,\mu$W$\else \,$\mu$\hbox{W}\fi}
\def\kms{\ifmmode $\,km\,s$^{-1}\else \,km\,s$^{-1}$\fi}
\def\kmsMpc{\ifmmode $\,\kms\,Mpc\mo$\else \,\kms\,Mpc\mo\fi}
\providecommand{\sorthelp}[1]{}

\newcommand{\nside}{\mathrm{N_{side}}}
\newcommand{\nlay}{N}
\newcommand{\alphaM}{\alpha_{\rm M}}
\newcommand{\fM}{f_{\rm M}}
\newcommand{\eqdef}{\,\hat{=}\,}
\newcommand{\hi}{\ensuremath{\mathsc {Hi}}}
\def\WMAP{\textit{WMAP}}
\newcommand{\stageA}{stage $a$}
\newcommand{\stageB}{stage $b$}
\newcommand{\stageAB}{stage $a$ and $b$}
\newcommand{\StageA}{Stage $a$}
\newcommand{\StageB}{Stage $b$}

\begin{document}

\title{Statistical simulations of the dust foreground  to cosmic microwave background polarization}
\author{F. Vansyngel\inst{1}, F. Boulanger\inst{1}, T. Ghosh\inst{1,2}, B. Wandelt\inst{3,4}, J. Aumont\inst{1}, A. Bracco\inst{1,5}, F. Levrier\inst{6}, P. G. Martin\inst{7}, L. Montier\inst{8}}
\institute{
Institut d'Astrophysique Spatiale, CNRS, Univ. Paris-Sud, Universit\'{e} Paris-Saclay, B\^{a}t. 121, 91405 Orsay cedex, France
\and
California Institute of Technology, Pasadena, California, U.S.A.
\and
Sorbonne Universit\'{e}s, UPMC Univ Paris 6 et CNRS, UMR 7095, Institut d’Astrophysique de Paris, 98 bis bd Arago, 75014 Paris, France
\and
Sorbonne Universit\'{e}s, Institut Lagrange de Paris (ILP), 98 bis Boulevard Arago, 75014 Paris, France
\and
Laboratoire AIM, IRFU/Service d'Astrophysique - CEA/DSM - CNRS - Universit\'{e} Paris Diderot, B\^{a}t. 709, CEA-Saclay, F-91191 Gif-sur-Yvette Cedex, France
\and
LERMA, Observatoire de Paris, PSL Research University, CNRS, Sorbonne Universités, UPMC Univ. Paris 06, Ecole normale supérieure, F-75005, Paris, France
\and
CITA, University of Toronto, 60 St. George St., Toronto, ON M5S 3H8, Canada
\and
CNRS, IRAP, 9 Av. colonel Roche, BP 44346, F-31028 Toulouse cedex 4, France
}

\abstract{
The characterization of the dust polarization foreground to the cosmic microwave background (CMB)  is a necessary step
toward the detection of the $B$-mode signal associated with primordial gravitational waves.
We present a method to simulate maps of polarized dust emission on the sphere that is similar  
to the approach used for CMB anisotropies. 
This method builds on the understanding of Galactic polarization stemming from the analysis of \Planck\ data. 
It relates the dust polarization sky to the structure of the Galactic magnetic field and its coupling with
interstellar matter and turbulence.
The Galactic magnetic field is modeled as a superposition of a mean uniform field and a Gaussian random (turbulent) component with a power-law power spectrum of exponent $\alphaM$. 
The integration along the line of sight carried out to compute Stokes maps 
is approximated by a sum over a small number of emitting layers with different realizations of the random component of the magnetic field. 
The model parameters are constrained to fit the power spectra of dust polarization $EE$, $BB,$ and $TE$ 
measured using \Planck\ data.  We find that the slopes of the $E$ and $B$ power spectra of dust 
polarization  are matched for $\alphaM = -2.5$, an exponent close to that measured for total dust intensity 
but larger than the Kolmogorov exponent -11/3. The model allows us 
to compute multiple realizations of the Stokes $Q$ and $U$ maps for different realizations of the random component of
the magnetic field, and to quantify the variance of dust polarization spectra for any given sky area outside of the Galactic plane. 
The simulations reproduce the scaling relation between the dust polarization power and the mean total dust intensity including the observed dispersion around the mean relation.
We also propose a method to carry out multifrequency simulations,
including the decorrelation measured recently by \Planck, using a given covariance matrix of the polarization maps.
These simulations are well suited to optimize component separation methods and to quantify the confidence with which the  
dust and CMB $B$-modes can be separated in present and future experiments. We also provide an astrophysical 
perspective on our phenomenological modeling of the dust polarization spectra.}

\titlerunning{Simulations of dust foreground to CMB polarization} 
\authorrunning{F. Vansyngel et al.} 
\maketitle


\section{Introduction}
An era of exponential expansion of the universe, dubbed cosmic inflation, has been proposed to explain why the universe 
is almost exactly Euclidean and nearly isotropic \citep{Guth81,Linde82}.
One generic prediction of this theoretical paradigm is the existence of a background of gravitational waves, which 
produces a distinct, curl-like, signature in the polarization of the cosmic microwave background (CMB), 
referred to as primordial $B$-mode polarization  \citep{Starobinsky79}. 
The detection of this signal would have  a deep impact  on  cosmology   and   fundamental physics,     
 motivating a number of experiments designed to measure the 
sky polarization at microwave frequencies. Current projects have achieved the sensitivity required to detect the CMB
$B$-mode signal predicted by the simplest models of inflation \citep{Abazajian15,Kamionkowski15}.
Yet, any detection has relied on the proper removal of much brighter Galactic foregrounds.

Thermal emission from aspherical dust grains aligned with respect to the 
Galactic magnetic field (GMF) is the dominant  polarized foreground for frequencies higher than about $70\,$GHz \citep{Dunkley09,planck2014-a12}. 
From the analysis of the \Planck\footnote{\Planck\ (\url{http://www.esa.int/Planck}) is a project of the European Space Agency (ESA) 
with instruments provided by two scientific consortia funded by ESA member states and led 
by Principal Investigators from France and Italy, telescope reflectors provided through a collaboration between 
ESA and a scientific consortium led and funded by Denmark, and additional contributions from NASA (USA).}
 $353\,$GHz polarization maps, we know that the 
primordial $B$-mode polarization of the CMB cannot be measured without subtracting  the foreground emission, even in the faintest dust-emitting regions  
at high Galactic latitude \citep[][hereafter PXXX]{planck2014-XXX}.
The observed correlation between the $B$-mode signal detected by BICEP2/Keck Array, on the one hand, and the \Planck\ dust maps, on the other hand, has confirmed this conclusion
\citep{pb2015}.  

To distinguish cosmological and Galactic foreground polarization signals, CMB experiments must rely on multifrequency observations. 
Component separation is a main challenge because the spatial structure of dust polarization is observed to vary with 
frequency \citep{planck2016-L}. This introduces two questions that
motivate our work. 
What design of CMB experiments and combination of ground-­based, balloon-­borne, and space observations
is best to achieve an optimal separation? How can confidence in the subtraction of foregrounds be quantified?
To provide quantitative answers, we must be able to simulate observations of the polarized sky
combining Galactic and CMB polarization. This paper presents a statistical model with a few parameters  
to simulate maps of dust polarization in a way similar to what is available for CMB anisotropies \citep{Seljak96}. 

The  analysis of the Wilkinson Microwave Anisotropy Probe (\WMAP) data and the preparation of the \Planck\ project 
motivated a series of models of the polarized synchrotron and thermal dust emission at microwave frequencies 
\citep{Page07,Miville08,Fauvet11,ODea12,Delabrouille13}. These early studies followed two distinct approaches. The first is to produce a sky that is as 
close as possible to the observed sky combining data templates and 
a spectral model. Prior to \Planck, for dust polarization this was performed using stellar polarization data by \citet{Page07}, and 
\WMAP\ observations of synchrotron polarization by \citet{Delabrouille13}. More recently, 
the simulations presented in \citet{planck2014-a14} use the \Planck\  $353\,$GHz data 
to model dust polarization. This first approach is limited by the signal-to-noise ratio of available data, which for dust polarization is low at high Galactic latitude even after smoothing to one degree angular resolution. The second approach is to 
simulate the polarization sky from a 3D model of the GMF and of the density structure of the interstellar medium (ISM), both its regular and turbulent components, 
as carried out by \citet{Miville08}, \citet{Fauvet11} and \citet{ODea12}. This method connects the modeling of the microwave polarized sky to 
broader efforts to model the GMF  \citep{Waelkens09,Jansson12,Planck2016-XLII}.

 \Planck\ polarization maps have been used to characterize the structure \citep{planck2014-XIX,planck2014-XX} 
and the spectral energy distribution (SED) of polarized thermal emission from Galactic dust \citep{planck2014-XXI,planck2014-XXII}. 
Several studies have established the connection between the structure of the magnetic field and matter 
\citep{Clark14,planck2014-XX,planck2014-XXXII,Martin15,Kalberla16}. 
The power spectra analysis presented in PXXX decomposes dust polarization into $E$ (gradient-like) and $B$ (curl-like)
modes \citep{Zaldarriaga01,Caldwell16}. This analysis led to two unexpected
results: a positive $TE$ correlation and a ratio of about
2 between the $E$ and $B$ dust powers over the
$\ell$ range 40 to 600.  
\citet{Clark15} and \citet{planck2015-XXXVIII} have showed that the
observed $TE$ correlation and asymmetry between $E-$ and $B$-mode 
power amplitudes for dust polarization  could be both accounted
for by  the preferred alignment  between the filamentary structure of the total intensity map
and the orientation of the magnetic field inferred from the
polarization angle. 

The work presented here makes use of  the model framework introduced 
in  \citet{planck2016-XLIV} (hereafter PXLIV). By analyzing \Planck\ dust polarization maps toward the southern Galactic cap, the part of the sky used for CMB 
observations from Antartica and Atacama, PXLIV related the large-scale patterns of the maps to the mean orientation of the 
magnetic field, and the scatter of the dust polarization angle and fraction ($\psi$ and $p$) to the amplitude of its turbulent component. 
In this paper, we extend their work to produce Stokes maps that fit dust polarization power spectra including the $TE$ correlation and 
the $TT/EE$ and $EE/BB$ power ratios at high and intermediate Galactic latitudes. 
In a companion paper \cite{Ghosh16}, the dust polarization of the southern sky region 
with the lowest dust column density is modeled using \hi\ observations
and astrophysical insight to constrain their model parameters. In essence, our approach is more mathematical but it allows us to model dust polarization over a larger fraction of the sky.
The two approaches are complementary and compared in this paper.
We also present a mathematical process to introduce spatial decorrelation  across 
microwave frequencies via the auto and cross spectra of dust polarization.
By doing this, we obtain a model to compute independent realizations 
of dust polarization sky maps at one or multiple frequencies with a few parameters adjusted to fit the statistical properties inferred from the analysis of the 
\Planck\ data away from the Galactic plane.  

The paper is organized as follows. 
Sects.~\ref{sec:model} and \ref{sec:fullsky} present the framework we use to model dust polarization in general terms. 
Our method is illustrated by producing simulated maps at $353\,$GHz presented in Sect.~\ref{sec:simulated_maps}. 
We show that these maps successfully match the statistical properties of dust polarization derived from the analysis 
of \Planck\  data (Sect.~\ref{sec:simspec}). 
One method to compute dust polarization maps at multiple frequencies is presented in Sect.~\ref{sec:multi-frequency}. We discuss the astrophysical implications of 
our work in Sect.~\ref{sec:astro}. The main results of the paper are summarized in Sect.~\ref{sec:summary}. 
Appendix~\ref{Appendix:A} details how the simulated maps used in this study are computed. Appendix~\ref{Appendix:B} shows how to compute the cross correlation between two frequency maps, when spectral differences about a mean SED may be parametrized with a spatially varying spectral index.

\section{Astrophysical framework}
\label{sec:model}

To model dust polarization we used the framework introduced by PXLIV, which we briefly describe here. 
We refer to PXLIV for a detailed presentation and discussion of the astrophysical motivation
and the simplifying assumptions of our modeling approach. 

The polarization of thermal dust emission results from the alignment of aspherical grains with respect to the GMF \citep{Stein66,Lee85,planck2014-XXI}. 
Within the hypothesis that grain polarization properties, including alignment, are homogeneous, the structure of the 
dust polarization sky reflects the structure of the magnetic field combined with that of matter. We assume that this 
hypothesis applies to the diffuse ISM where radiative torques provide a viable mechanism to align grains efficiently \citep[][]{Dolginov76,Andersson15,Hoang16}.

To compute the Stokes parameters $I$, $Q$, and $U$ describing the linearly polarized  thermal dust emission, 
we start from the integral equations in Sect.~3.2 and Appendix B of \citet{planck2014-XX} for optically thin emission at frequency $\nu$, i.e.,
\begin{align}
\label{eq:intIQU} 
\StokesI(\nu)=\int S(\nu) \,\left[1-p_0\left(\cos^2\polangsky-\frac{2}{3}\right)\right]\mathrm{d}\tau_\nu; \nonumber \\
\StokesQ(\nu)=\int p_0\,S(\nu) \,\cos\left(2\phi\right)\cos^2\polangsky\,\mathrm{d}\tau_\nu; \\
\StokesU(\nu)=\int p_0\,S(\nu) \,\sin\left(2\phi\right)\cos^2\polangsky\,\mathrm{d}\tau_\nu. \nonumber 
\end{align}
where $S(\nu)$ is the source function, $\tau_\nu$ the optical depth, 
$p_0$ a parameter related to dust polarization properties (the grain cross sections and the degree of alignment with the magnetic field), $\gamma $  
the angle that the local magnetic field makes with the plane of the sky, and $\phi $ the local polarization angle (see Fig.~14 in \citet{planck2014-XX}). 

As in PXLIV,  the integration along the line of sight is approximated by a sum 
over a finite number $\nlay$ of layers. This sum is written as 
\begin{align}
\label{eq:sumIQU} 
& I (\nu) = \sum^{\nlay}_{i=1}  S_i(\nu) \, \left[1-p_0\left(\cos^2\gamma_i-\frac{2}{3}\right)\right] ;  \nonumber \\
& Q(\nu)  = \sum^{\nlay}_{i=1} p_0  \, S_i(\nu) \, {\rm cos}(2\phi_i) \,  {\rm cos}^2\gamma_i ;   \\ 
& U(\nu)  =  \sum^{\nlay}_{i=1} p_0 \, S_i(\nu)  \, {\rm sin}(2\phi_i) \,  {\rm cos}^2\gamma_i ; \nonumber 
\end{align}
where $S_i (\nu)$ is the integral of the source function over layer  $i$, 
and $\gamma_i$ and $\phi_i$ define the magnetic field orientation within each layer.   
As discussed in PXLIV,
the layers are a  phenomenological means to model the density structure of the  interstellar matter and the correlation 
length of the GMF.
This approach accounts for both signatures of the  turbulent magnetic field component in Galactic polarization maps: the depolarization 
resulting from the integration along the line of sight of emission with varying polarization orientations, and the 
scale invariant structure of the polarization maps across the sky reflecting the power spectrum of the turbulent component of the magnetic field \citep{Cho02,Houde09}. 
It  overcomes the difficulty 
of generating realizations of the turbulent component of the magnetic field in three dimensions over the celestial sphere. \cite{Ghosh16} uses \hi\ data to associate the layers with different phases of the ISM, each of which provide a different intensity map. On the contrary, in the simulations presented in this paper, like in PXLIV, the term $S_i(\nu)$ in Eqs.~\ref{eq:sumIQU} is a sky map  
assumed to be the same in each layer, i.e., it is independent of the index $i$. Thus we do not address the question of the physical meaning of the layers.

Through the angles $\gamma_i$ and $\phi_i$,  the model  relates 
the dust polarization to the structure of the GMF.  
The magnetic field $\vec{B} $ is expressed as the sum of its mean (ordered), $\vec{B}_0$, 
and turbulent (random), $\vec{B}_{\rm t}$, components,
\begin{equation}
\label{eq:vecB} 
\vec{B}  = \vec{B}_0 + \vec{B}_{\rm t} = |\vec{B}_0| \,(\vec{\hat{B}}_0 + \fM \, \vec{\hat{B}}_{\rm t})
,\end{equation}
where $\vec{\hat{B}}_0$ and $\vec{\hat{B}}_{\rm t}$ are unit vectors in the directions of $\vec{B}_0$ and $\vec{B}_{\rm t}$, and $\fM$
a model parameter that sets the relative strength of the random component of the field. 
To simulate dust as a foreground to  the CMB we need a description of the GMF within the solar neighborhood. 
We follow PXLIV in assuming that $\vec{B}_0$ has a fixed orientation in all layers. 
We ignore the structure of the GMF on galaxy-wide scales because the dust emission arises mainly from a thin 
disk with a relatively small scale height and
we are interested in modeling dust polarization away from the Galactic plane. 
This scale height is not measured directly in the solar neighborhood but modeling of the dust emission from the Milky Way
indicates that it is $\sim 200\,$pc at the solar distance from the Galactic center \citep{Drimmel01}.  

Each component of the vector field $\vec{\hat{B}}_{\rm t}$ in 3D, in each layer, is obtained from
independent Gaussian realizations of a power-law power spectrum, which is written as
\begin{equation}
\label{eq:specBt} 
C_\ell \propto \ell^{\alphaM}{\rm  ~for ~} \ell \ge 2.
\end{equation}
 Our modeling of $\vec{\hat{B}}_{\rm t}$ is continuous over the celestial sphere and uncorrelated between layers.
The coherence of the GMF orientation along the line of sight comes from the mean field
and is controlled by the parameter $\fM$. 

The model has six parameters: the Galactic longitude and latitude $l_0$ and $b_0$ defining the orientation of  \vec{\hat{B_0}}, the factor
$\fM$, the number of layers $\nlay$,  the spectral exponent $\alphaM$, and the effective polarization fraction of the dust emission $p_0$. 
The PXLIV authors used the same model to analyze the dust polarization measured by \Planck\ at 353\,GHz over the southern Galactic cap (Galactic latitude $b < -60^\circ$). 
They determined $l_0 = 70\deg\pm5\deg$ and $b_0 = 24\deg\pm5\deg$ by fitting the large-scale pattern observed in the Stokes
$Q$ and $U$ maps, and $\fM= 0.9\pm 0.1$, $\nlay = 7\pm 2 $ and $p_0 = 26\pm 3$\% by fitting the distribution function (one-point statistics) 
of $p^2$, the square of the dust polarization fraction $p$, and of the polarization angle $\psi$,  computed after removal of the regular pattern
from the ordered component of the GMF. 

Hereafter we label the Stokes maps computed from Eqs.~\ref{eq:sumIQU}  as $I_a,Q_a$, and $U_a$.
At this \stageA, the power spectra of the model maps have equal $EE$ and $BB$ power, and no $TE$ correlation at $\ell \gtrsim 30$.
This follows from the fact that our modeling does not include the alignment observed between the filamentary structure of the diffuse ISM and the GMF orientation.
Some $TE$ correlation is present at low $\ell$  because the mean GMF orientation is close to being within the Galactic disk,
and we take into account the latitude dependence of the total dust intensity. 
In the next section, we explain how we modify the spherical harmonic decomposition of the \stageA\ maps to 
introduce the $TE$ correlation and the $E$-$B$ asymmetry, matching the \Planck\ dust polarization power spectra in PXXX.

\section{Introducing TE correlation and E-B asymmetry}
\label{sec:fullsky}

Our aim is to simulate maps that match  given observables based on dust angular power spectra, namely the $TE$ correlation,  
$TT$/$EE$  and $EE$/$BB$  ratios, and the $BB$ spectrum without altering the statistics of $p$ and $\psi$ of the Stokes maps from \stageA .
We describe a generic process to construct such a set of Stokes maps $(I_b,Q_b,U_b)$, later referred  to as \stageB\ maps. 
The process can be applied on a full, or a masked, sky.

We start with the Stokes maps $(I_a,Q_a,U_a)$ obtained as described in Sect.~\ref{sec:model}. We compute  the spherical harmonic coefficients of the \stageB\ maps from those of the \stageA\ maps as follows:
\begin{equation}
\begin{dcases}
\label{eq:blm}
b_{\ell m}^T &= t a_{\ell m}^T \\
b_{\ell m}^E &= p_0(a_{\ell m}^E/p_0 + \rho a_{\ell m}^T) \\
b_{\ell m}^B &= p_0(f a_{\ell m}^B/p_0)
\end{dcases} \, .
\end{equation}
where $a_{\ell m}^{X}$ and $b_{\ell m}^{X}$ denote the coefficients of the $X=T,E,B$ harmonic decomposition of \stageAB\ maps, respectively. 
The parameter $\rho$ introduces the $TE$ correlation and the factor $f$ the $E$-$B$ asymmetry. 
The parameter $t$ is a scaling factor for the intensity part and $p_0$ is the polarization parameter introduced in Eqs.~\ref{eq:intIQU}. 
These parameters control the amplitude of the $TT$, $EE$, $BB,$ and $TE$ power spectra of \stageB\ maps.
We note that $Q_a$ and $U_a$ scale linearly with $p_0$ and thus that the two ratios $a_{\ell m}^E/p_0$ and $a_{\ell m}^B/p_0$ in Eqs.~\ref{eq:blm} are independent of $p_0$.

At this \stageB, our modeling of the random component of the magnetic field is anisotropic, which is a fundamental characteristic of magnetohydrodynamical turbulence \citep{2012ApJ...747....5L,Brandenburg13}.
 The factors $\rho$ and $f$ introduce anisotropy in two ways. First, the $T$ map, which is added to the polarization part through the parameter $\rho$, has a filamentary structure and thus is anisotropic. This amounts to adding an extra polarization layer that is perfectly aligned with the filamentary structure of the matter and is similar to what is carried out by \cite{Ghosh16} for their cold neutral medium map. Second, the factor $f$ breaks the symmetry between $E$ and $B$, whereas the power is expected to be equally distributed between $E$ and $B$ modes in the case of isotropic turbulence \citep{Caldwell16}. Through the parameter $f$, the random component of \vec{B} is anisotropic in all layers and everywhere on the sky, unlike in \cite{Ghosh16} where anisotropy is introduced in only one layer.

In the simplest case, $\rho,f,t,$ and $p_0$ are constants over the whole multipole range and in the most general case they are functions of $\ell$ and $m$. We find that 
the statistics of  $p$ and $\psi$ found using the \stageA\ maps are lost at \stageB\
if $f \neq 1$ or $\rho \neq 0$ for very low multipoles. 
Thus, we look for a solution where the parameters $t$ and $p_0$ are constants but $f$ and $\rho$ depend on $\ell$ and tend toward 1 and 0 for very low $\ell$ values, respectively.

The power spectra of \stageB\ maps are noted  $C_\ell^{XY} $ with $X, Y=T, E, B$ and use the quantity ${\cal D}_\ell^{XY}\equiv\ell(\ell+1) \, C_\ell^{XY}/(2\pi)$.
The $t$, $p_0$, $\rho,$ and $f$ coefficients in Eqs.~\ref{eq:blm} are chosen such that the power spectra of \stageB\ maps match a given set of averaged ratios as follows:
\begin{align}
\label{eq:ratios}
& R_{TT}\equiv {\cal E}\left[{\cal D}_\ell^{TT}/{\cal D}_\ell^{EE}\right],  \nonumber \\
& R_{TE}\equiv {\cal E}\left[{\cal D}_\ell^{TE}/{\cal D}_\ell^{EE}\right],     \\
& R_{BB}\equiv {\cal E}\left[{\cal D}_\ell^{BB}/{\cal D}_\ell^{EE}\right] , \nonumber
\end{align}
where ${\cal E}\left[\cdot\right]$ is a given averaging process over multipoles.
The absolute scaling is performed by matching the amplitude of one power spectrum. For this purpose, we use the $BB$ spectrum because the main motivation of the simulations is to 
produce polarized dust skies for component separation of $B$-modes. Thus,  to Eqs.~\ref{eq:ratios} 
we add the fourth constraint
\begin{equation}
\label{eq:NB}
N_B \eqdef \left(p_0f\right)^2 \, ,
\end{equation}
where $N_B$ is an overall factor that scales the $BB$ power spectrum of \stageA\ maps divided by $p_0$ to the desired amplitude. The four parameters $t$, $p_0$, $\rho,$ and $f$ can be derived analytically from the four input parameters $R_{TT}$, $R_{TE}$, $R_{BB}$, and $N_B$.
One can choose any values for $R_{TT}$, $R_{TE}$, $R_{BB}$, and $N_B$, as long as the normalization is positive and the ratios respect the condition $R_{TT} > R_{TE}^2$ forced by the positive definiteness of the power spectra covariance.

We construct the $b_{\ell m}^T$, $b_{\ell m}^E$, and $b_{\ell m}^B$
 according to their definitions in Eqs.~\ref{eq:blm}.  The final product is a triplet of Stokes maps $(I_b,Q_b,U_b)$ 
that have the desired two-point statistics.

\section{Simulated maps}
\label{sec:simulated_maps}

To illustrate our method, we apply the formalism presented in the previous sections and simulate dust polarization maps that fit the \Planck\ power spectra.
The input values for $R_{TT}$, $R_{TE}$, and $R_{BB}$ in Eqs.~\ref{eq:ratios} are derived from  \Planck\ data  (Sect.~\ref{sec:Planckspectra}).
We introduce the simulated maps in Sect.~\ref{sec:simus}. The method used to compute these maps is detailed in Appendix~\ref{Appendix:A}.

\subsection{\Planck\ power spectra}
\label{sec:Planckspectra}

The $EE$, $BB$, $TE,$ and $TB$ angular power spectra of dust polarization were measured using the \Planck\ maps at $353\,$GHz on the six large regions at high and intermediate Galactic latitude defined in PXXX. The effective sky fraction $f_{sky}$, after a $5^\circ$ (FWHM) apodization, ranges from $f_{sky}=24\%$ to $f_{sky}=72\%$. The regions are labeled LRxx, with xx the sky fraction in percent.

The $EE$ and $BB$ spectra reported in PXXX are well fitted by power laws with exponents $\alpha_{EE,BB}^{data} = -2.42 \pm 0.02 $, with no systematic dependence on the sky region. 
The amplitudes of the spectra at a reference multipole $\ell_0 = 80$, $A^{EE,data}$, were measured from power-law fits over the range $40< \ell < 600$ with an index fixed to its mean value of $-2.42$. 
These amplitudes are observed to increase with the mean total dust intensity in the mask, $I_{dust}$, following the law $A^{XX,data}\propto \left(I_{dust}\right)^{1.90 \pm 0.02}$ ($X=E,B$).
We combine  the amplitude $A^{EE,data}$
and the $EE$ to $BB$ ratios listed in Table~1 of PXXX for their LR33 mask 
to compute the amplitude $A^{BB,data}$ of the ${\cal D}_{\ell}^{BB,data}$ spectrum at $\ell=80$. 

The values of the $R_{TT}$ and $R_{TE}$ ratios are not listed in Table~1 of PXXX. To determine these values, we combine the fit to the $EE$ spectrum from PXXX, 
the TE spectrum plotted in Fig.~B.1 of PXXX, and the TT spectrum
we computed using the Planck dust map at 353 GHz obtained by~\cite{planck2016-XLVIII} after separation from the cosmic infrared background (CIB) anisotropies.
The ${\cal C}_\ell$ data points and error bars of the TE spectrum were provided to us by the contact author of PXXX. The spectra are binned between $\ell=40$ and $\ell=600$ with $\Delta\ell=20$ and the binned spectra are noted ${\cal C}_b^{XY,data}$ ($XY=TT,TE$).
We compute the ratios $R_{XY}$ by comparing the measured power spectra ${\cal C}_b^{XY,data}$  with the power-law fit to the $EE$ spectrum ${\cal C}_b^{EE,data}$, minimizing the following chi-squared:
\begin{equation}
\chi^2(R) = \sum_b \left( {\cal C}_b^{XY,data}-R \; {\cal C}_b^{EE,data} \right)^2 / (\sigma_b^{XY,data})^2\, ,
\end{equation}
where
$\sigma_b^{XY,data}$ is the standard deviation error on ${\cal C}_b^{XY,data}$ output from the Xpol power spectrum estimator\footnote{Xpol is an algorithm for power spectrum estimation that is an extension to polarization of the Xspect method \citep{Xspect}}.

The values we use as input for the simulations are gathered in Table~\ref{tab:ratios}.

\begin{table}
\centering
\caption{Input values for the simulations.}
\label{tab:ratios}
\begin{tabular}{r||c|c|c|c|c}
$f_{sky}$ & $R_{TT}$ & $R_{TE}$ & $R_{BB}$ & $\alpha_{BB}^{data}$ & $A^{BB,data}$ \\
   & & & & & $\muK_{\text{CMB}}^2$ \\
\hline
\hline
33\% & 44.2$\pm$3& 2.5$\pm$0.2 & 0.48$\pm$0.03 & -2.37$\pm$0.12 & 24.5$\pm$1.7  \\
\end{tabular}
\end{table}

\begin{table}
\centering
\caption{Values of the parameters $t$, $p_0$, $\rho,$ and $f$ corresponding to the ratio and normalization values of Table~\ref{tab:ratios} and to our fiducial set of values for $\nlay$, $\fM$, and $\alphaM$.}\label{tab:tpaf}
\begin{tabular}{c|c|c|c}
$t$ & $p_0$ & $\rho$ & $f$ \\
\hline
\hline
1.01$\pm$0.15 & 0.22$\pm$0.05 & 0.25$\pm$0.03 & 0.75$\pm$0.02
\end{tabular}
\end{table}

\subsection{Simulated maps used in this study}
\label{sec:simus}

Here and in Appendix~\ref{Appendix:A}, we introduce the simulated maps and describe how we produce them. 

We have analyzed the simulated maps over a larger sky area 
than in PXLIV. We have not, however, attempted to fit the PXLIV model of the mean field to the \Planck\ data over a larger region.
In particular, the adopted mean field direction is given by the same Galactic coordinates $(l_0,b_0)=(70^\circ,24^\circ)$. Although this specific choice affects the $Q_a$ and $U_a$ maps, it has no critical impact on the statistical results presented in the paper.
Our fiducial set of values for $\nlay$, $\fM$, and $\alphaM$ is  4, 0.9, and $-2.5$, respectively.
To quantify the impact of these parameters on the model power spectra, we computed simulated maps for  several combinations  around the fiducial values within the constraints set by PXLIV. For $\nlay$ we considered 
two values 4 and 7, and for $\fM$ the range 0.7 to 1.0. We explored a range of values of $\alphaM$ from $-3.4$ to $-2.2$.

The method we followed to construct the \stageAB\ maps is described in Sects.~\ref{subsec:modelA} and \ref{subsec:modelB}. We produced our simulations at an angular resolution of $30 \arcmin$ on a HEALPix\footnote{\url{http://healpix.sourceforge.net}} \citep{gorski2005} grid with resolution parameter $\nside=256$. Although the parameter $p_0$ was computed at \stageB, we needed an initial guess in order to compute the total intensity map of \stageA\ maps (see Eqs.~\ref{eq:sumIQU} for $I(\nu)$). Based on PXLIV, we took $p_0= 0.25$.  We used this value to
compute \stageB\ maps from $Q_a/p_0$ and $U_a/p_0$ that do not depend much on $p_0$ (Sect.~\ref{subsec:modelA} ).

The parameters $t,p_0,\rho,$ and $f$ used to construct \stageB\ maps were determined by the ratios 
$R_{TT}$, $R_{TE}$, and $R_{BB}$ and the amplitude of the $BB$ spectrum (Sect.~\ref{subsec:modelB}). We used the $BB$ amplitude and the ratio values computed on the LR33 mask (Table~\ref{tab:ratios}). 
The corresponding values of the \stageB\ parameters $t$, $p_0$, $\rho$ and $f$ are listed in Table~\ref{tab:tpaf}  for
our fiducial set of values for $\nlay$, $\fM$ and $\alphaM$. 
The value of $p_0$, $0.22 \pm 0.05$  agrees with that derived by PXLIV from their data fit, which we used to compute the \stageA\ maps. 
Thus, it is not necessary to iterate the process. The scaling factor $t$ of the Stokes $I$ map is found to be unity within uncertainties.   

Because the \stageA\ maps have a high intensity contrast, the conversion from pixel space to spherical harmonic space induces leakage of power from the Galactic plane to high latitudes. 
In order to avoid this artifact, the brightest part of the Galactic plane must be masked before performing the transformation. 
The \Planck\ collaboration provides eight Galactic masks for general purposes.
They are derived from the 353 GHz intensity map by gradually thresholding the intensity after having subtracted the CMB. These masks are then apodized with a 2 degree Gaussian kernel and cover respectively 15, 33, 51, 62, 72, 81, 91, and 95\% of the sky\footnote{These masks are available on the Planck Legacy Archive as  \texttt{HFI\_Mask\_GalPlane-apo2\_2048\_R2.00.fits} and described in the Planck Explanatory Supplement 2015 accessible at the web page \url{https://wiki.cosmos.esa.int/planckpla2015/index.php/Frequency_Maps\#Galactic_plane_masks}}. 
The precise choice of the mask  is not critical. We chose the mask corresponding to $f_{sky}=80\%$, which 
discards low Galactic latitude areas where our model with a uniform mean orientation of the field does not apply. The unmasked 
region is large enough to encompass all regions outside the Galactic plane that are relevant 
for CMB analyses.

\begin{figure}
\centerline{\includegraphics[scale=.9]{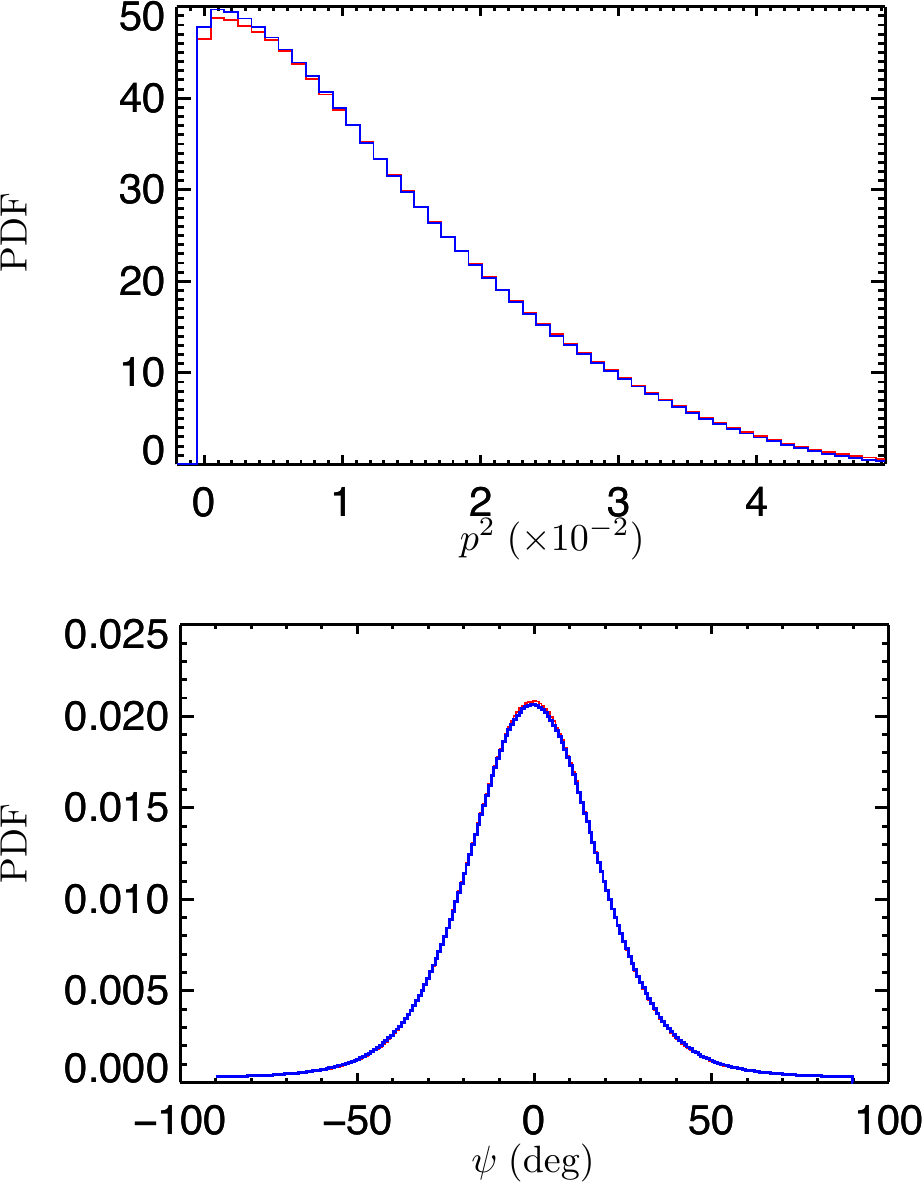}}
\caption{Probability distribution functions of $p^2$ and $\psi$ (top and bottom plots) for \stageA\ and \stageB\ maps (red and blue histograms). The maps were computed using the fiducial values of $\alphaM$, $\fM$, and $\nlay$ and the corresponding parameters $t$, $p_0$, $\rho,$ and $f$ introduced in Sect.~\ref{sec:simus}. The distributions are computed on the southern 
Galactic polar cap ($b \le -60^\circ$) as in PXLIV. The very close match between the corresponding histograms shows that the 
inclusion of the $TE$ correlation and the $E$-$B$ asymmetry does not alter the one-point statistics of the simulated maps.}\label{fig:histoppsi}
\end{figure}

As mentioned in Sect.~\ref{sec:fullsky}, extending the $E$-$B$ asymmetry down to very low multipoles changes the one-point statistics of fraction and angle of polarization.
To prevent this artefact, we introduce the $E$-$B$ asymmetry and the $TE$ correlation smoothly from low multipoles. In practice, the parameters $\rho$ and $f$ are functions of $\ell$ as follows:
\begin{equation}
\begin{cases}
\rho(\ell) &= \rho w(\ell) \\
f(\ell) &= 1-(1-f)w(\ell)
\end{cases} \, .
\end{equation}
Here $w(\ell)$ is a window function going smoothly from 0 to 1 around multipole $\ell_c$ and is defined as follows:
\begin{equation}
 w(\ell)=
 \begin{cases}
 0 & \text{if } \ell \leq \ell_c - \delta\ell/2 \\
 \left(1-\text{sin}\left(\frac{\ell_c-\ell}{\delta\ell}\;\pi\right)\right)\,/\,2 & \text{if } \ell_c - \delta\ell/2 < \ell < \ell_c + \delta\ell/2 \\
 1 & \text{if }  \ell_c + \delta\ell/2 \leqslant \ell
 \end{cases}
,\end{equation}
where we set $\ell_c = 30$ and $\delta\ell = 30$. After this modification, the $E$-$B$ power ratio tends to 1 for $\ell < \ell_c$   in agreement
with  the $EE$ and $BB$  \Planck\ $353\,$GHz power spectra presented in Fig.~20 of \citet{planck2016-XLVI} at $\ell < 30$.
Figure~\ref{fig:histoppsi} shows that the distributions (one-point statistics) of $p$ and $\psi$ computed around the southern Galactic pole of the \stageAB\ maps are very similar.

\begin{figure*}
\centerline{\includegraphics[scale=1]{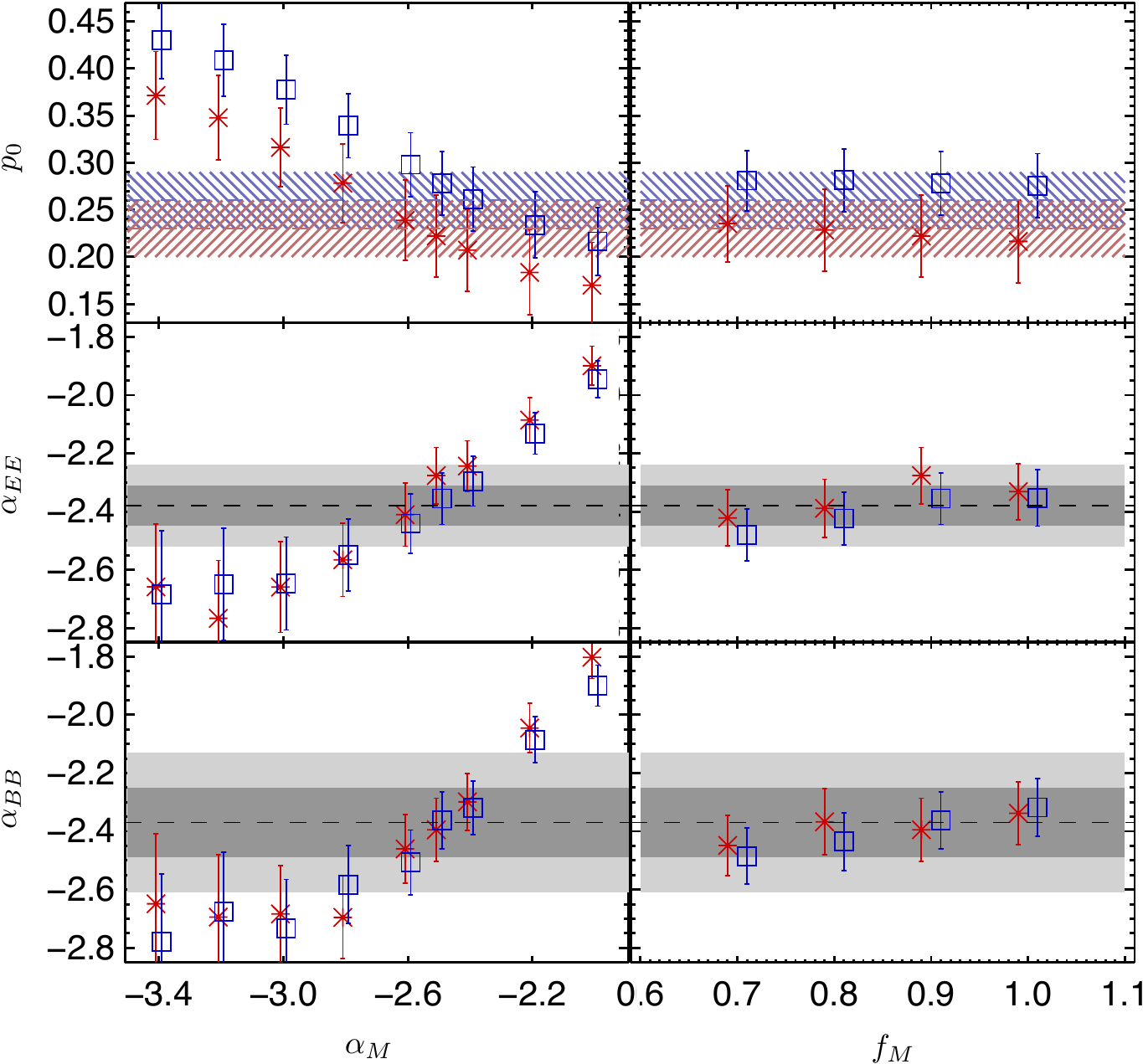}}
\caption{Parameter $p_0$ (top) and slopes of the $EE$ (middle) and $BB$ (bottom) power spectra of \stageB\ maps vs. the slope $\alphaM$ of the power spectrum of the turbulent component of the magnetic field (left) and vs. the relative strength of the turbulence $\fM$ (right). In the left plots $\fM=0.8$ and in the right plots $\alphaM = -2.5$. Red stars (blue squares) represent results for $\nlay=4$ ($\nlay=7$). The abscissae of the two sets of points are slightly shifted from their original values for a better visibility. The values observed in the data are represented by a gray shaded region for $\alpha_{EE}$ and $\alpha_{BB}$ (a dashed line for the mean, dark, and light gray for the 1- and 2$\sigma$ uncertainties) and by a hatched regions for $p_0$ (a dashed horizontal line for the mean, and a red 45$^\circ$ (resp. blue -45$^\circ$) hatched region for 1$\sigma$ uncertainty for $\nlay = 4$ (resp. 7)).
}\label{fig:fmavEE} 
\end{figure*}

\section{Model power spectra}
\label{sec:simspec}

In this section, we show that our simulated \stageB\ maps reproduce the \Planck\ $EE$, $BB,$ and $TE$ dust spectra constraining the exponent  $\alphaM$ of the magnetic field power spectrum
(Sect.~\ref{subsec:matching}),  provide the statistical variance of the dust polarization power in a given $\ell$ bin  (Sect.~\ref{subsec:distribpow}), 
and match the observed scaling between the spectra amplitude and the mean dust total intensity for both large and small sky regions (Sect.~\ref{subsec:powvar}).
 
\subsection{Matching \Planck\ power spectra}
\label{subsec:matching}

To compare our model results directly with the analysis of the \Planck\ data in PXXX, we compute power spectra of the simulated maps over the LR33 mask.
The power spectra are computed using the PolSpice estimator \citep{PolSpice} that corrects for multipole-to-multipole coupling due to the masking. We checked that we obtain very similar results when the spectra are computed with the Xpol estimator.

For both values $\nlay=4$ and 7, we vary the parameters $\fM$ and $\alphaM$ as follows. First, we keep $\fM$ fixed to 0.9 and let $\alphaM$ vary from $-3.4$ to $-2.0$ in steps of 0.2 with the addition of $-2.5$, then we keep  fixed $\alphaM$  to $-2.5$ and let $\fM$ vary from 0.7 to 1 in steps of 0.1. For each set of parameters, we compute a sample of 1000 realizations with the procedure described in Sect.~\ref{sec:simus} and Appendix~\ref{Appendix:A}. 
The power spectra of \stageB\ maps are binned from $\ell=60$ to 200 with a bin width of $\Delta\ell = 20$. We fit the model $A^{XX}\left(\ell/\ell_0\right)^{\alpha_{XX}+2}$ ($X=E$ or $B$, $\ell_0=80$) to the sample mean spectrum ${\cal D}_\ell^{XX}$. The weights used in the fit are the entries of the sample covariance matrix.
For each pair of $(\fM,\alphaM)$ values, we can derive the mean and covariance of $(A^{XX}, \alpha_{XX})$ from the fit.

Fig.~\ref{fig:fmavEE} shows the changes in the parameter $p_0$ and the spectral indices $\alpha_{EE}$ and $\alpha_{BB}$ when varying either  $\fM$ or 
$\alphaM$, for $\nlay=4$ and $7$. The points are the sample means of the parameters $p_0$, $\alpha_{EE}$, and $\alpha_{BB}$ and the error bars represent the sample standard deviation. The results are compared to the data values reported in PXLIV and in PXXX. In PXLIV, the authors constrain the value of $p_0$ with one-point statistics of the $p^2$ and $\psi$ around the south pole
at a fixed number of layers $\nlay$ (see middle plot of Fig.~10 of PXLIV).
Over  the range of values we consider, $p_0$, $\alpha_{EE}$, and $\alpha_{BB}$  are mostly sensitive to $\alphaM$. 
The comparison  of the power spectra between simulations and data does not constraint $\fM$ nor $\nlay$. 
The parameter $\fM$ affects both the dispersion of $\psi$ and $p$ through depolarization along the line of sight (PXLIV). 
These two effects modify the variance of the dust polarization in opposite directions.
The fact that the parameter $p_0$ is independent of $\fM$ (see top right panel of Fig.~\ref{fig:fmavEE}) suggests that they compensate each other over the 
range of values we are considering.

The measured values of $\alpha_{EE}$ and $\alpha_{BB}$ constrain $\alphaM$ to be $-2.5$ within about 0.1.
For  steeper $\vec{B}_{\rm t}$ spectra ($\alphaM \leq -2.8$), 
$\alpha_{EE}$ and $\alpha_{BB}$ are roughly constant with mean values lower than the observed values.
 In this regime, turbulence is not significant over the $\ell$ range used in this analysis. The dust total intensity map and the changing orientation with respect to the line of sight of the 
 mean magnetic field dominate the variance of the polarized maps. For $\alphaM \geq  -2.6$, $\alpha_{EE}$ and $\alpha_{BB}$  are roughly equal to
$\alphaM$ within a small positive offset of about 0.1. In other words, 
the exponents of the dust polarization spectra  reproduce the exponent of the magnetic field power spectrum.

The parameter $p_0$ may also be used to constrain $\alphaM$.
If the  $p_0$ values from PXLIV for $\nlay=4$ and 7 hold for the LR33 region, 
we find that the model fit constrains $\alphaM$ to be $-2.5$ within an uncertainty of about 0.1 (top left panel of Fig.~\ref{fig:fmavEE}).
The systematic dependence of $p_0$ with $\alphaM$ follows from dispersion of the $\vec{B}_{\rm t}$ orientation on angular scales corresponding to multipoles $\ell > 40$.
For a given $\fM$, this dispersion decreases as the power spectrum of $\vec{B}_{\rm t}$ steepens (i.e., toward low values of $\alphaM$). 
Hence, the observed amplitude of the $BB$ spectrum is matched for increasing values of $p_0$ when $\alphaM$ decreases.

\begin{figure}
\centerline{\includegraphics[scale=1]{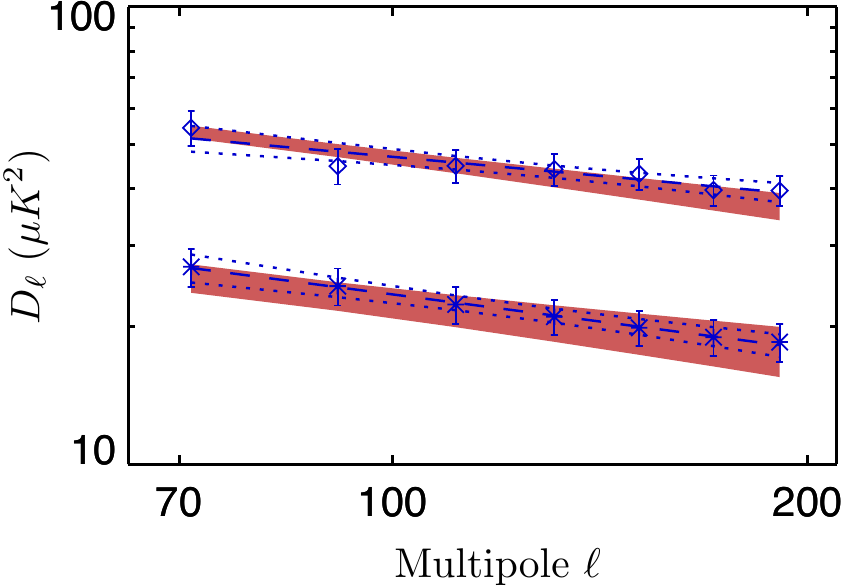}}
\caption{$EE$ (top curve, diamond symbols) and $BB$ (bottom curve, star symbols) power spectra of the simulated maps and their fits for the LR33 sky region. The diamonds and the stars represent the mean value computed over 1000 realizations. The 1$\sigma$ error bars are derived from the sample standard deviation of the power in each $\ell$ bin. The blue dashed lines represent the fits to the mean spectra and the blue dotted lines the 1$\sigma$ error on the fits. The red shade areas represent the power-law fit and  the 1$\sigma$ errors to the \Planck\ data reported in PIPXXX for the LR33 region.}\label{fig:PSref}
\end{figure}

Fig.~\ref{fig:PSref} shows the $EE$ and $BB$ power spectra for our fiducial values of  $\alphaM$, $\fM$, and $N$. The points represent the mean value computed over 1000 realizations. The errors are derived from the sample variance of the power in each $\ell$ bin. The fit from the analysis of PXXX and its 1$\sigma$ error are overplotted. The simulations are able to reproduce the $EE$ and $BB$ dust power spectra.
The asymmetry parameter $f$ has a value smaller than unity. The factor $f^2=0.55$ is close to the value of $R_{BB}=0.48$ (Table~\ref{tab:ratios}). Within this model, unlike for that of
\citet{Ghosh16},  the $TE$ correlation accounts for only a small part of the $E$-$B$ asymmetry.

\begin{figure}
\centerline{\includegraphics[scale=1]{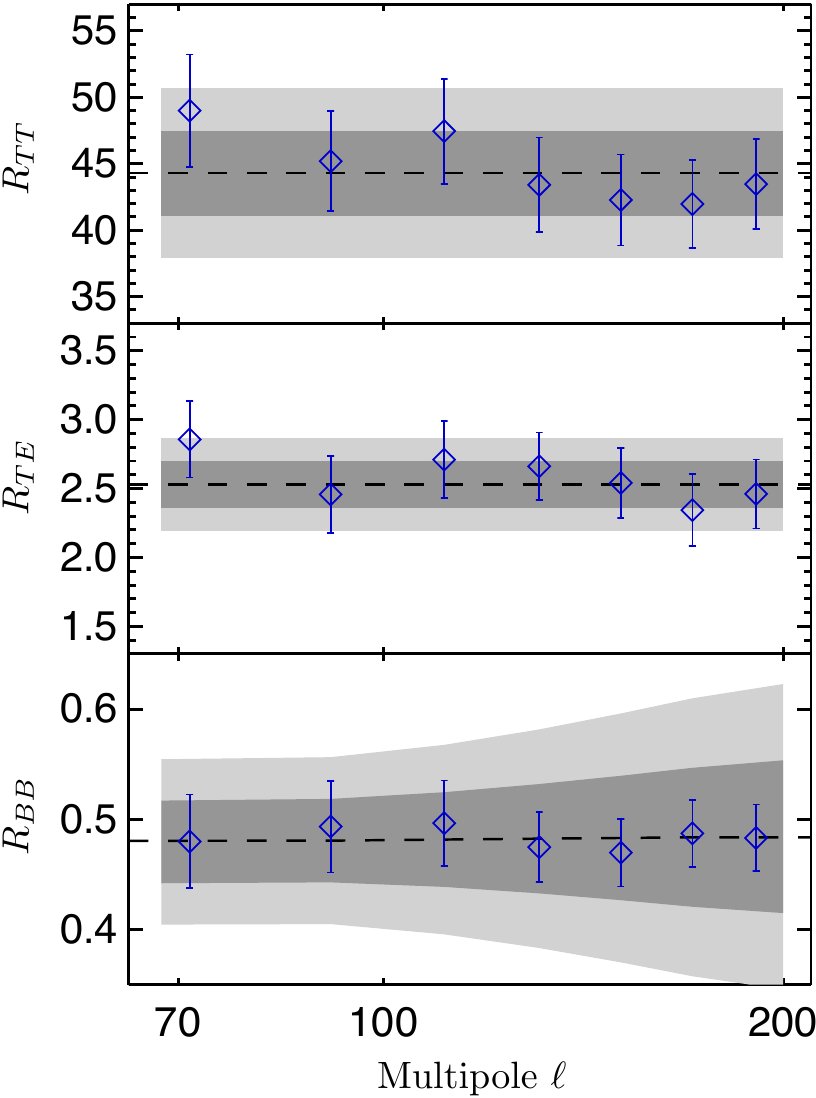}}
\caption{Three ratios $R_{TT}$, $R_{TE}$, and $R_{BB}$ computed on the simulated maps for the LR33 sky region. 
The blue diamonds are the mean ratios for each $\ell$ bin computed over 1000 realizations. For the $R_{TT}$ and $R_{TE}$ ratios, the dashed line and the light and dark gray regions represent the input value and the 1- and 2$\sigma$ errors on the input value, respectively. For the $R_{BB}$ ratio, the dashed line, the light and dark gray regions represent, respectively, the ratio between the fits of the $BB$ and $EE$ data spectra from PXXX and their 1- and 2$\sigma$ uncertainties.}\label{fig:Rref}
\end{figure}

Fig.~\ref{fig:Rref} shows ratios between the different power spectra of the simulated maps. Each point represents the sample mean of the 1000 ratios ${\cal D}_\ell^{TT}/{\cal D}_\ell^{EE}$, ${\cal D}_\ell^{TE}/{\cal D}_\ell^{EE}$ and ${\cal D}_\ell^{BB}/{\cal D}_\ell^{EE}$ of each bin and the error bars represent the sample standard deviation. 
For comparison, we plot the input values and uncertainties of the $R_{TT}$, $R_{TE}$, and $R_{BB}$ ratios.
The ratios computed on the simulated maps are consistent with the input values, as expected because
the maps were constructed in such a way that their power spectra respect that covariance structure.

\subsection{Statistics of the power spectrum amplitudes}
\label{subsec:distribpow}

Our simulations allow us to compute the dispersion of the dust $BB$ power within a given $\ell$ bin. Although 
the dust maps are computed from Gaussian realizations of the turbulent field, the 
various processes involved in the computation might make them non-Gaussian. 
For example, we do not expect the distribution of the power at multipole $\ell$ to tend to a Gaussian distribution for $\ell\rightarrow\infty$ as quickly as it would for a Gaussian random field. For the same reason, the variance of the distribution of the power for a given $\ell$  is not necessarily the cosmic variance.

\begin{figure}
\centerline{\includegraphics[scale=1]{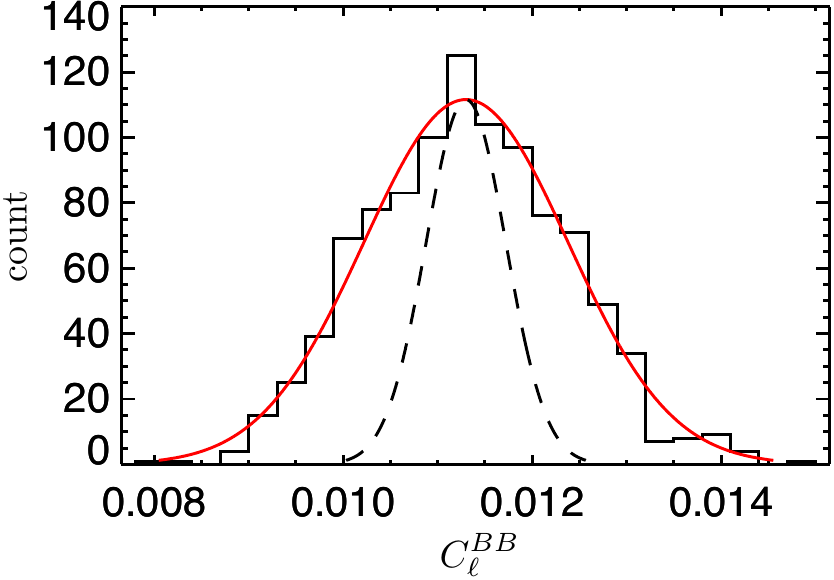}}
\caption{Distribution of the power per bin computed on simulated maps is significantly broader than the cosmic variance. The solid red line represents a Gaussian fit to the distribution in the multipole 
bin $\ell=110$ with a width  $\Delta\ell=20$ (black histogram). The dashed line represents the distribution expected for a Gaussian random field in that same bin.}\label{fig:disrat}
\end{figure}

Fig.~\ref{fig:disrat} shows the distribution of the power within one multipole bin around $\ell=110$ with a bin width of $\Delta\ell = 20$. The power spectra were computed for the LR33 region\ for which the covered sky is roughly equally distributed around the north and south Galactic poles. The figure also presents 
a Gaussian fit to the histogram and the expected cosmic variance for the same bin if the maps were drawn from a Gaussian random field on the sphere. 
The actual dispersion is a few times larger than the cosmic standard deviation. This effect might be due to the non-stationarity of the intensity map. The LR33 region includes 
some bright structures in dust total intensity. These localized  structures are likely to be the explanation for the enhanced dispersion in the simulations. If this is the right interpretation,
the enhancement must apply to the true sky because we are using the \Planck\ total dust intensity map in our model. 

In addition to the spread, we looked at the shape of the PDF of the power per bin. We made 10000 of our dust simulations and 10000 Gaussian random simulations. The power spectrum used to produce the Gaussian realizations is the sample mean power spectrum of the 10000 dust simulations. For the two cases, we computed the power spectra, binned them with a width of $\Delta\ell=20,$ and fitted a Gaussian function to the sample distribution. In both cases, the dust simulations and the Gaussian realizations, we see the same difference between the PDF of the power per bin and the Gaussian fit. We concluded that the shape of distribution of the power per bin of our simulations is very similar to that of a Gaussian random field.


\subsection{Power variations over the sky}
\label{subsec:powvar}

We now show that the simulations reproduce the \Planck\ power spectra for the high latitude sky in general, not just for the specific sky region LR33 used as input.
First, we compute the spectra of the simulated maps for the five other LRxx sky regions from PXXX. 
Second, as in PXXX, we compute the spectra for smaller sky  patches at high Galactic latitude with $f_{sky} =1\%$. We compare the amplitudes of the 
simulations spectra with the \Planck\ results.

The analysis of the simulations on the six  regions provides six sample mean power spectra and their sample variances. The power spectra on each region are computed and are fitted in the same way as 
described in Sect.~\ref{subsec:matching}. In Table~\ref{tab:fits}, we gather the results of the fits together with the corresponding \Planck\ values collected from Table~1 of PXXX for comparison. Error bars on the data measurements are smaller than those of these noiseless simulations because the error bars on the simulation spectra contain the variance from multiple random realizations of the GMF that does not affect the data.

While the simulations are constructed such that they match the data on one  particular sky region (LR33), Table~\ref{tab:fits} shows that they also agree with the  data on the other five regions within 
a small difference, which we  comment on below.
The spectra amplitudes at $\ell=80$ increase with the sky fraction faster than what PXXX reported for the \Planck\ data. 
This slight difference may arise from the fact that we assumed a fixed value of $\nlay$ independent of the dust total intensity and Galactic latitude. In models 
of stellar polarization data at low Galactic latitudes and in molecular clouds, \cite{Jones92,Myers91} assumed that $\nlay$ scales linearly with the dust column density. While 
their model hypothesis would not work for the diffuse ISM, we could consider variations in $\nlay$. 
Alternatively, the slight difference in scaling could come from another simplifying assumption of the method, as we ignore the variation of the mean GMF orientation with distance from the Sun. It will be possible to modify our model to test these two ideas but this is beyond the scope of the present paper.

\begin{table*}
\centering
\begin{tabular}{r||c|c|c|c|c|c}
$f_{\mathrm{sky}}^{\mathrm{eff}}$& 0.24   &              0.33              &              0.42                &           0.53                &           0.63                 &       0.72  \\
\hline
\hline
 $A^{EE}\;(\mu K^2)$  & 30.6$\pm$2.3 & 50.4$\pm$2.8 & 95.2$\pm$4.9 & 157.6$\pm$7.7 & 261$\pm$12 & 419$\pm$20 \\
 $\alpha_{EE}$ & -2.37$\pm$0.13 & -2.307$\pm$0.099 & -2.438$\pm$0.092 & -2.334$\pm$0.087 & -2.385$\pm$0.084 & -2.413$\pm$0.088 \\
 $r_{A\alpha,E}$ & -0.85 & -0.84 & -0.83 & -0.83 & -0.82 & -0.81 \\
 $\chi^2 (N_{dof}=5)$  &  0.1 & 0.7 & 0.4 & 0.8 & 1.4 & 1.0 \\
 \hline
 $A^{BB}\;(\mu K^2)$  & 14.3$\pm$1.1 & 24.7$\pm$1.5 & 44.9$\pm$2.4 & 74.0$\pm$4.0 & 123.6$\pm$6.4 & 196$\pm$11 \\
 $\alpha_{BB}$ & -2.35$\pm$0.14 & -2.33$\pm$0.11 & -2.439$\pm$0.099 & -2.313$\pm$0.098 & -2.332$\pm$0.094 & -2.40$\pm$0.10 \\
 $r_{A\alpha,B}$ & -0.82 & -0.82 & -0.81 & -0.81 & -0.81 & -0.81 \\
 $\chi^2 (N_{dof}=5)$  &  0.6 & 0.1 & 0.7 & 0.9 & 0.5 & 1.4 \\
 \hline
 \hline
 $A^{EE,data}\;(\mu K^2)$         &   37.5$\pm$1.6      &  51.0$\pm$1.6          &         78.6$\pm$1.7      &       124.2$\pm$1.9  & 197.1$\pm$2.3          &    328.0$\pm$2.8         \\
 $\alpha_{EE}^{data}$ &   -2.40$\pm$0.09  & -2.38$\pm$0.07        &   -2.34$\pm$0.04         &  -2.36$\pm$0.03     &  -2.42$\pm$0.02       &  -2.43$\pm$0.02      \\
 \hline
 $A^{BB,data}\;(\mu K^2)$ & 18.4$\pm$1.7 & 24.5$\pm$1.7 & 41.7$\pm$1.8 & 67.1$\pm$2.7 & 104.5$\pm$2.3 & 173.8$\pm$3.6          \\
 $\alpha_{BB}^{data}$ &   -2.29$\pm$0.15 &   -2.37$\pm$0.12         &  -2.46$\pm$0.07       &    -2.43$\pm$0.05  &    -2.44$\pm$0.03        & -2.46$\pm$0.02     \\
 \end{tabular}
\caption{Results of power-law fits to the power spectra computed on simulated dust maps for the six Galactic regions from PXXX. The quantities $r_{A\alpha,X}$ and $\chi^2$ are the correlation between $A^{XX}$ and $\alpha_{XX}$ and the value of the $\chi^2$ at the fit values, respectively. Values from the \Planck\ data taken from PXXX are given for comparison.}\label{tab:fits}
\end{table*}

\begin{figure}
\centerline{\includegraphics[scale=1]{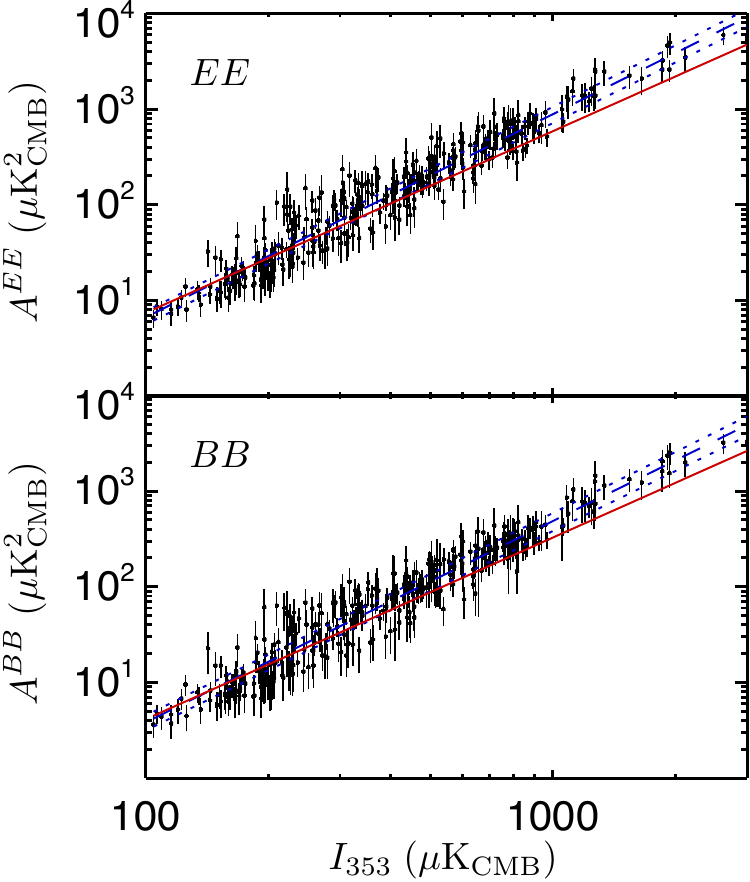}}
\caption{Amplitudes of the power spectra ($A^{EE}$ and $A^{BB}$) plotted vs. the mean total dust intensity at $353\,$GHz computed on each $1 \%$ sky region for 100 realizations. Each vertical black line represents the sample mean and sample standard deviation of one of the $400\, $deg$^2$ patch. The blue dashed and dotted lines represent the power-law fit with the 1$\sigma$ uncertainty of the simulation results. For comparison the red line is the same fit to the \Planck\ data for the same set of sky patches. }\label{fig:emplaw}
\end{figure}

For the analysis on the 1\% sky patches, we perform the same procedure as in PXXX to derive the empirical law between the amplitude at $\ell=80$ of the power spectra and the total intensity. Fig.~\ref{fig:emplaw} shows the amplitudes of the $EE$ and $BB$ spectra as a function of the mean intensity of each patch. We realized 100 simulations and each vertical black line represents the sample mean and sample dispersion amplitude of
one $400\, $deg$^2$ patch. The empirical law derived from a linear fit in the $\log(I_{353})-\log(A^{XX})$ space is overplotted. From this fit, we find a slope value of 2.15$\pm$0.03 for the $EE$ spectrum and of 2.09$\pm$0.03 for the $BB$ spectrum. The values of the slopes are slightly larger than 1.9$\pm$0.02, which is the value that was measured on the \Planck\ data. This difference for the patches is similar to that observed for the large sky regions, where the amplitude of the power spectra increases with $f_{sky}$ slightly faster in the simulation than in the data (see Table~\ref{tab:fits}).

In PXXX, the authors found that the cosmic variance and their measurement uncertainties were not large enough to account for the dispersion around the fit of Fig.~\ref{fig:emplaw}. For our simulations, the spread of the distribution of the power in a given $\ell$ bin shown in Fig.~\ref{fig:disrat} can explain the scatter observed around the fit of Fig.~\ref{fig:emplaw}, which is comparable to that seen in the data. As detailed in Sect.~\ref{subsec:distribpow}, the scatter in the model comes mainly from the turbulent component of the magnetic field. In particular, we checked that the spread around the line fit is correlated with the mean polarization fraction, which depends on the mean orientation of the magnetic field over a given sky patch.


\section{Multifrequency simulations}
\label{sec:multi-frequency}

So far we have discussed ways to simulate structures on the sky at a single reference frequency. Component separation methods for CMB experiments rely on multifrequency data. A common approach to multifrequency simulations is to simulate  the sky structure and the  SED separately. The SED can be simulated using templates or analytical forms relying on a set of parameters, such as a modified blackbody law.
The simulated sky map at a given frequency is then extrapolated to other frequencies. This method could also be applied to our simulations, but it does not permit us to control the  
decorrelation between maps at different frequencies in harmonic space, which is a characteristic crucial for  component separation as discussed in \citet{planck2016-L}. Indeed, the decorrelation has an impact on the relative weights between the principal foreground modes.
Here we present a method for multifrequency simulations constrained to match a given set of auto- and cross-power spectra.

\subsection{Method}

We follow a procedure close to that commonly used to compute pseudo-random Gaussian vectors with a desired covariance from  vectors with unit covariance. We realize as many simulated dust polarization maps as the desired number of frequencies and rearrange them to form a new set of maps such that the covariance structure of the latter is exactly as wanted.

To build a set of $N_f$ maps at frequencies $\left\lbrace \nu_i, i=1\dots N_f \right\rbrace$, we proceed as follows:
\begin{enumerate}
\item Simulate $N_f$ single-frequency maps obtained as described in Sect.~\ref{sec:fullsky}, whose polarization spherical harmonic coefficients are gathered in a $2N_f$-dimension (E and B for $N_f$ maps) vector $x_{\ell m}$ for each pair $(\ell,m)$.
\item Compute the auto- and cross-power spectra of the maps and gather them in a matrix $\Sigma_\ell$, which is $2N_f\times 2N_f$ at each multipole $\ell$.
\item Specify a covariance structure of the maps over the range of $N_f$ frequencies in the form of a $2N_f\times 2N_f$ matrix $\cal{C}_\ell$ for each multipole $\ell$. 
\item For each multipole, $\ell$, compute the Cholesky decomposition of $\Sigma_\ell$ and $\cal{C}_\ell$, i.e.,
\begin{eqnarray}
\Sigma_\ell &=& L_\ell L_\ell^\dagger \, , \\
\cal{C}_\ell &=& M_\ell M_\ell^\dagger \, ,\end{eqnarray}
where the superscript $\dagger$ denotes the transposition.
\item For each pair ($\ell$,$m$), construct the $2N_f$-dimension vector
\begin{equation}
y_{\ell m} = M_\ell L_\ell^{-1} x_{\ell m} \, .
\end{equation}
\end{enumerate}
It can be easily verified that the set of maps whose spherical harmonics coefficients are gathered in $y_{\ell m}$ has exactly the expected auto- and cross-spectra.

\begin{figure}
\centerline{\includegraphics[scale=1]{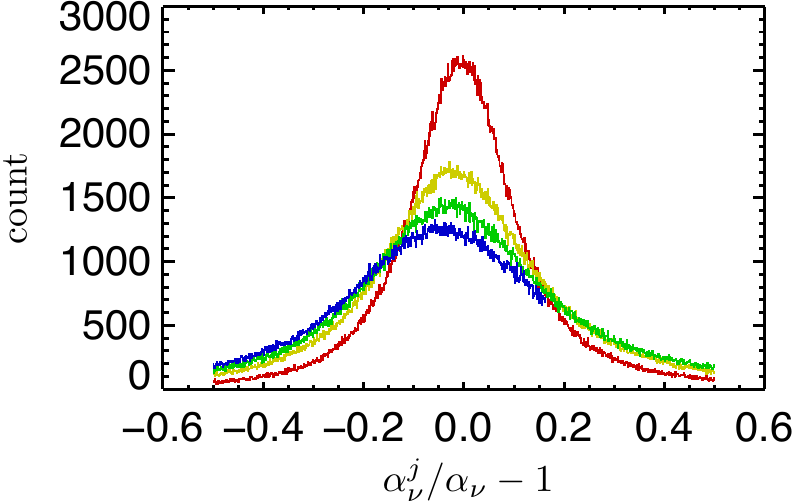}}
\caption{Distribution of the relative difference to the mean SED, normalized to 1 at 353 GHz in the 217 GHz (red, the narrower), 143 GHz (yellow), 100 GHz (green), and 70 GHz (blue, the wider) maps.}\label{fig:SED}
\end{figure}

\subsection{Results}

We applied the procedure to produce a multifrequency set of maps $(I(\nu,j),Q(\nu,j),U(\nu,j))$ where $\nu=70,100,143,217,353$ GHz and $j=1\dots N_p$ is the pixel index. For both $EE$ and $BB$, the diagonal of the imposed covariance is $C_\ell^{\nu\times\nu}=(\nu/\nu_0)^{2\, \beta}(B_\nu(T_0)/B_{\nu_0}(T_0))^2 C_\ell^{\nu_0\times\nu_0}$, where $\nu_0=353$ GHz, $T_0=19.6\,$K, $\beta=1.6$ \citep{planck2014-XXII} and $C_\ell^{\nu_0\times\nu_0}$ is the power spectrum of simulations at frequency $\nu_0$. The SED-independent correlation ratio  $R_\ell=C_\ell^{\nu_1\times\nu_2}/\sqrt{C_\ell^{\nu_1\times\nu_1}C_\ell^{\nu_2\times\nu_2}}$ between two frequencies $\nu_1$ and $\nu_2$ is set to 1 below $\ell=30$ and set by the following equation above $\ell=30$:
\begin{equation}
R_\ell = \exp\left\lbrace -\frac12\sigma^2 \left[\log\left(\frac{\nu_1}{\nu_2}\right)\right]^2\right\rbrace \, .
\end{equation}
This dependence applies if the variations of the SED can be parametrized with  a spatially varying spectral index (Appendix \ref{Appendix:B}). The parameter $\sigma$ is set in such a way that the correlation between the 353 and 217\,GHz channels is 0.9 within the range of values measured on \textit{Planck} data \citep[][]{planck2016-L}.
We then construct the SED map $\alpha_\nu^j$ from
\begin{equation}
\alpha_\nu^j = \frac{\sqrt{Q(\nu,j)^2+U(\nu,j)^2}}{\sqrt{Q(\nu_0,j)^2+U(\nu_0,j)^2}}
\end{equation}
and compute the mean SED $\alpha_\nu$ from
\begin{equation}
\alpha_\nu = \left(\prod_j \alpha_\nu^j\right)^{1/N_p} \, .
\end{equation}

In Fig.~\ref{fig:SED}, we plot the distribution of $\alpha_\nu^j / \alpha_\nu-1$ for each  $\nu=70,100,143,$ and $217$ GHz. As expected, the distribution widens with the separation between $\nu$ and $\nu_0$ because the correlation coefficient $R_\ell$ decreases. The correlation between the normalized SED of the same four frequencies is given by
\renewcommand{\kbldelim}{(}
\renewcommand{\kbrdelim}{)}
\[\kbordermatrix{
 & 217 & 143 & 100 & 70 \\
217 &  1      &     0.91  &    0.80 &     0.74 \\
143 & 0.91  &    1         &  0.94   &   0.86 \\
100 & 0.80   &   0.94    &  1         &  0.96 \\
 70  & 0.74    &  0.86     & 0.96     &  1
}\, .
\]

This matrix gives an estimation of the coherence of the normalized SED through frequencies.
We do not control the way the SED of a given sky pixel varies with respect to the mean SED because  we model the decorrelation in harmonic space statistically.

\section{Astrophysical perspective}
\label{sec:astro}

Our paper has so far focused on our contribution to component separation for CMB data analysis. 
We presented a phenomenological model that can be  used to simulate 
dust polarization maps, which statistically match \Planck\ observations and are noise-free.  In this section,
we discuss what we learn about the GMF in the local interstellar medium from the modeling of 
the dust polarization power spectra.  
We examine our model results from this astrophysical perspective. We also compare our results with those of a companion paper
\citet{Ghosh16}, which uses \hi\ data to account for the multiphase structure of the diffuse ISM.  
In Sect.~\ref{subsec:model_fit}, we briefly review \Planck\
power spectra of dust polarization and our model fit.  
In Sects.~\ref{subsec:alpha_GMF}  and  \ref{subsec:correlation_GMF}, we discuss the power spectrum of the GMF 
and its correlation with matter. 

\begin{figure}
\centerline{\includegraphics[scale=1]{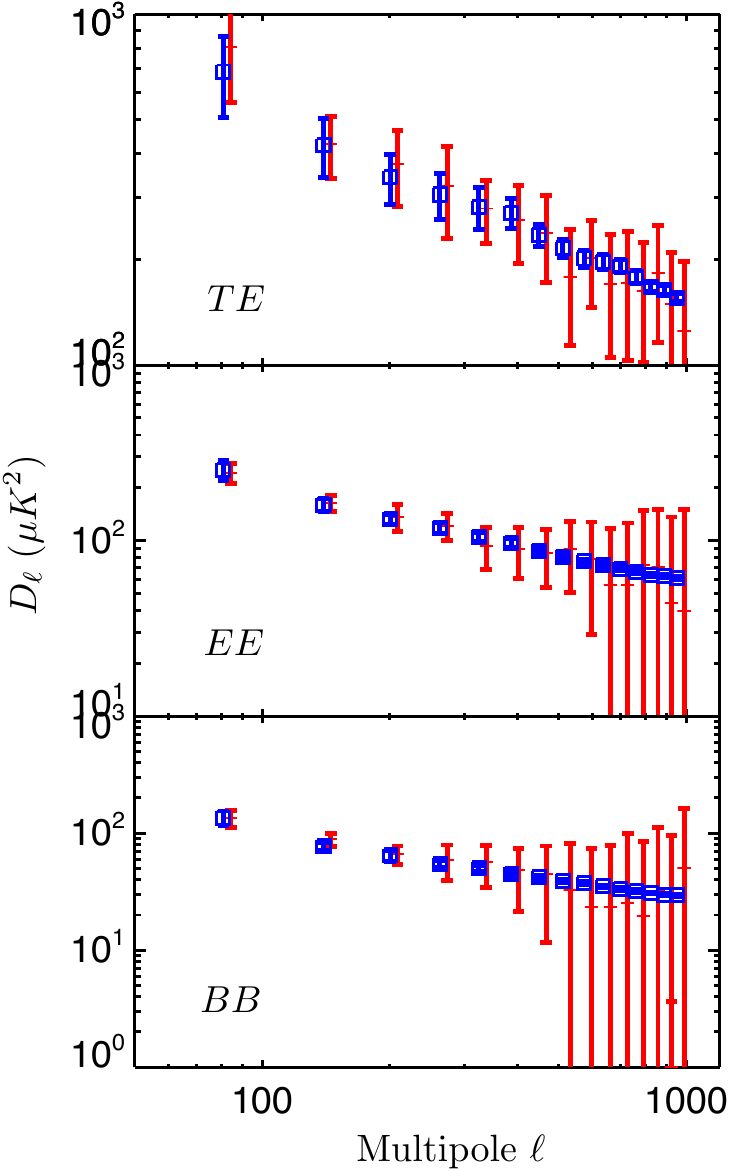}}
\caption{High resolution power spectra on the LR63 region; simulation vs. data. From top to bottom: $TE$, $EE,$ and $BB$ power spectra of the \Planck\ $353\,$GHz, CMB-corrected maps  (black), and one high resolution (${\rm N_{side}}=2048$, FWHM = $10 \arcmin$) realization of the model (red).}\label{fig:highres}
\end{figure}

\subsection{Model fit of the dust polarization spectra}
\label{subsec:model_fit}

We computed one simulation at an angular resolution of $10\arcmin$ ($\ell \simeq 1000$)
to illustrate the model fit of the \Planck\ data over a wider range of  multipoles than in
Sect.~\ref{sec:simspec}. The $TE$, $EE,$ and $BB$ spectra are presented in Fig.~\ref{fig:highres}. 
The data spectra are cross-spectra computed over the LR63 region using the two half-mission maps at $353\,$GHz
of \Planck\  \citep{planck2014-a01,planck2014-a09} after subtraction of the corresponding half-mission {\it SMICA} CMB maps \citep{planck2014-a11}. 
The simulation is built for our  fiducial parameters of the turbulence. The values of the four parameters $(t,p_0,\rho,f)$  were determined for this sky region and this specific realization 
to be $(1.01, 0.22, 0.20,0.74)$.  

The spectra in Fig.~\ref{fig:highres} are consistent with a  single spectral exponent  over multipoles $40 \le \ell \le 1000$. At $\ell > 1000$, the \Planck\ spectra are dominated by the noise variance. 
At $\ell < 40$, the spectra we computed with the publicly available maps are not reliable due to uncorrected systematics. 
The $EE$ and $BB$ \Planck\ 353\,GHz spectra computed down to $\ell =2 $ after systematics corrections are presented in \citet{planck2016-XLVI}. 
These spectra shown in their Fig.~20 indicate a flattening at $\ell < 20$, which is more pronounced for $EE$ than for $BB$; the $E$ to $B$ 
power ratio goes from about 2 to 1 toward low multipoles.  

An effective distance to the emitting dust is necessary to convert multipoles into physical scales.
Over the high Galactic latitude region LR63,  we estimate the  distance of the emitting dust to be in the range 100-$200\,$pc. 
This estimate is constrained by the distance to the edge of the local bubble \citep{Lallement14} and the scale height of the dust 
emission, $200\,$pc at the solar Galacto-centric radius from the model of  \citet{Drimmel01}. For the upper value of this distance range, 
the multipole range 40-1000 corresponds to linear scales from 0.5  to $15\,$pc. 

\subsection{Galactic magnetic field power spectrum}
\label{subsec:alpha_GMF}

Three of the model parameters we use -- $\fM$, $\nlay$ and $p_0$ -- were constrained 
in PXLIV. Within these constraints, we find that our model fits the dust polarization power spectra 
for a spectral exponent of the $\vec{\hat{B}}_{\rm t}$ power spectrum
$\alphaM = -2.5 \pm 0.1$ (Sect.~\ref{subsec:matching}). Within the quoted uncertainty, 
this value matches the spectral exponent of $-2.42 \pm 0.02$ of the \Planck\ dust that is 
measured over the same range of multipoles on the $EE$ and $BB$ $353\,$GHz \Planck\ spectra.
Thus, a main conclusion of our modeling is  that the exponent of the dust polarization spectra is that of the $\vec{\hat{B}}_{\rm t}$ spectrum.
The same conclusion is reached by \citet{Ghosh16} for a distinct modeling of the polarization layers.
This conclusion holds within the common framework of these two models and the corresponding assumptions.

The spectral exponent $\alphaM$ we derive from the data fit is significantly larger than 
the Kolmogorov value of $-11/3$ that is the common reference in interstellar turbulence \citep{Brandenburg13}, which is observed to apply to 
the electron density over a huge range of physical scales \citep{Armstron95,Cherpurnov10}. 
A similar difference has been reported for the GMF spectrum derived over a similar range of scales 
from the analysis of synchrotron emission  \citep[e.g.,][]{Iacobelli13} and of Faraday rotation measures \citep{Oppermann12}. 
As discussed theoretically for synchrotron emission by  \citet{Chepurnov98} and \citet{Cho02}, a shallower slope is expected for 
$\ell $ multipoles approaching $\pi \, {{L_{\rm max}}\over{L_{\rm out}}}$, where $L_{\rm max}$ is the length of the emitting layer along the line of sight 
and $L_{\rm out}$ the outer scale of turbulence. Two given lines of sight  
cross independent turbulent cells when their separation angle approaches the angle $\sim {{L_{\rm out}}\over{L_{\rm max}}}$.  It is only for smaller separation angles  that 
the power spectrum of the emission reflects that of the magnetic field. This explanation put forward for synchrotron emission 
and Faraday rotation in earlier studies could apply to our analysis of dust polarization too. The flattening 
observed at $\ell < 20$ in the spectra presented by \citet{planck2016-XLVI} supports this interpretation, but the   
\Planck\ data do not have the sensitivity to fully test it by checking whether the dust polarization spectra steepen at $\ell > 1000$.
Alternatively, the exponent of the GMF spectra might follow from the correlation of the magnetic field with interstellar matter. 
Indeed, \citet{Ghosh16} find an exponent of $-2.4$ for the $E$ map
they computed assuming a perfect alignment between the magnetic field and filamentary structure of their cold neutral medium \hi\ map.

\subsection{Correlation between matter and the GMF}
\label{subsec:correlation_GMF}

In this section we relate the structure of the GMF to that of the gas density in the diffuse ISM. The two are expected to be correlated to the extent that the magnetic field is frozen in matter. We note that this assumption might not hold everywhere \citep{2013Natur.497..466E}. The dust total intensity at $353\,$GHz is a tracer of interstellar matter within some limitations characterized in a number of studies \citep[e.g.,][]{planck2013-XVII,planck2013-p06b}, which are not a main concern for this discussion. 
The spectrum of the GMF we find is close to that measured for the dust total intensity. 
Over the same $\ell$ range, \cite{planck2016-XLVIII} report an exponent of $-2.7$ for the $TT$ spectrum of their $353\,$GHz map corrected for CIB anisotropies, 
and \citet{Ghosh16}  report a value of -2.6 for their total dust intensity map built from \hi\ data. 

Dust polarization data have been used to quantify the alignment of the magnetic field orientation with the filamentary structure of the 
diffuse ISM \citep{Clark14,planck2014-XXXII}. This is a striking facet of the correlation between matter and the GMF, which 
creates $TE$ correlation and thereby $E$-$B$ power asymmetry \citep{Clark15,planck2015-XXXVIII}. 
\citet{Ghosh16} presented a model of dust polarization where this correlation between matter and the GMF applies to one single polarization layer that is associated with
the cold neutral medium as traced by narrow \hi\ spectral lines. In their model that layer accounts for both
the $TE$ correlation and the $E$-$B$ asymmetry measured over the sky region with the lowest dust column density in the southern sky they analyzed.  
In our model,  the $TE$ correlation is introduced by adding one dust emission layer, where polarization is only in $E$-modes and is fully correlated to the $T$ map. This corresponds to the additive term proportional to the $\rho$ parameter in the second equation in Eqs.~\ref{eq:blm}.
The dust filamentary structures are present in all layers and the polarization results from the addition of the signals.   
We checked on the simulated images that this process introduces a
preferred alignment between the filamentary structure of the $T$ map
and the magnetic field orientation inferred from the polarization
angle, but this alignment is not as tight as that reported by
\citet{planck2015-XXXVIII} from their analysis of the most conspicuous filaments at high galactic latitudes in the \Planck\ data.
 This difference comes from the fact that we use the same intensity map for each layer. \citet{planck2015-XXXVIII}  shows that 
the filamentary structure of the cold neutral medium has a main contribution to the $E$-$B$ asymmetry but it does not  exclude a  
significant contribution related to the generic anisotropy of MHD turbulence, as suggested by \citet{Caldwell16}. We stress here that our modeling of the $E$-$B$ power asymmetry is mathematical. It does not constrain its physical origin.  
In this respect, our model is a framework that we are using to match the data statistically, but without a predictive power for astrophysics.

\section{Conclusion}
\label{sec:summary}

We introduced a process to simulate dust polarization maps, which may be used to statistically assess component separation methods in CMB data analysis. 
We detailed the simulation of dust polarization maps at one frequency before we introduced  
a mathematical means to produce maps at several frequencies and matched a given set of auto- and cross-spectra.
Our method and the main results obtained by analyzing the simulated maps are summarized here.

Our approach builds on earlier studies, i.e., the analysis of \Planck\
dust polarization data  and the model framework from PXLIV, which
relate the dust polarization sky to the structure of the GMF and interstellar matter.
The structure of interstellar matter is the dust total intensity map from \Planck.  
The GMF is modeled as a superposition of a mean uniform field and a Gaussian random (turbulent) component with a power-law power spectrum of exponent $\alphaM$. 
The integration along the line of sight performed to compute the Stokes maps 
is approximated by a sum over a small number of emitting layers with different realizations of the random GMF component. 
The mean field orientation, the amplitude of the random GMF component with respect to the mean component, the spectral exponent $\alphaM$, 
and the number of polarization layers are parameters common 
to the model from PXLIV. To  match the power spectra of dust polarization measured with the \Planck\ data, we add two main parameters ($\rho$ and $f$)
that introduce mathematically the $TE$ correlation and $E$-$B$ power asymmetry.
They are determined by fitting the Planck $353\,$GHz power spectra for $\ell > 40$ 
on one sky region at high Galactic latitude, LR33 from PXXX. 

The model allows us to compute multiple realizations of the Stokes $Q$ and $U$ maps for different realizations of the random component of
the magnetic field and to quantify the dispersion of dust polarization spectra for any given sky area away from the Galactic plane. 
The simulations reproduce the scaling laws between the dust
polarization power and the mean total dust intensity from \Planck,\ including the observed 
dispersion around the mean relation. 

This paper discusses what we learn about the GMF in the local interstellar medium from the modeling of 
the dust polarization power spectra.  We find that the slopes of the $EE$ and $BB$ power spectra of dust 
polarization measured by Planck are matched for $\alphaM = -2.5 \pm 0.1$.  
As in \citet{Ghosh16}, we find that, for our model, the exponent of the spectrum of $\vec{\hat{B}}_{\rm t}$
is very close to that of the dust polarization spectra. 
This exponent is larger than the Kolmogorov value of $-11/3$ but close to that measured for matter ($-2.7$), over the same region and range of multipoles ($\ell = 40-1000$),  
using the \Planck\ dust total intensity at 353\,GHz as a tracer. 
Our model does not allow us to comment on the origin of the $TE$
correlation and $E$-$B$ asymmetry.  

It would be possible to extend the model we presented in several ways, which might lead to fruitful explorations.  
To fit  dust polarization spectra down to the very low multipoles relevant for measuring $E$ and $B$-mode CMB polarization 
associated with the universe reionization, we might need 
to account for the injection scale of turbulence. Phenomenologically, this could be carried out by 
introducing a low-$\ell$ cutoff in the power spectrum of the magnetic field in Eq.~\ref{eq:specBt}.

Further model changes could also provide a better match to the data, in particular toward low Galactic latitudes.  
We have used a constant orientation for the mean GMF. A 3D model of the density structure of the 
Galactic ISM can be used to assign distances to the shells, and, thereby, to take into account the 3D structure of 
the large-scale magnetic field, as in, for example, \citet{Fauvet11} and \citet{Planck2016-XLII}.
In this case the intensity maps will differ for each layer and the effective number of layers could be allowed to vary with, for example, Galactic latitude or dust column density. 
In such a model, it would be possible  to introduce, for each layer, the correlation between matter and the GMF and distinct dust SEDs. This method of introducing the decorrelation of dust polarization maps with frequency might in essence better 
represent  the line-of-sight averaging of polarization data  \citep[][]{Tassis15,planck2016-L} than the mathematical means proposed here.  
Finally, our paper focuses on dust polarization but a similar approach could be applied to produce maps of synchrotron polarization 
that match the observed correlation with dust polarization 
\citep{planck2014-XXII,Choi15}.

\begin{acknowledgements}
The research leading to these results has received funding from the European
Research Council under the European Union's Seventh Framework Programme
(FP7/2007-2013) / ERC grant agreement No.~267934.
\end{acknowledgements}

\bibliographystyle{aa}
\bibliography{../Planck_bib,../FB_bib}

\addcontentsline{toc}{section}{Appendix}
\appendix

\section{Implementation of the method}
\label{Appendix:A}

In this appendix, we explain how we compute the dust polarization maps used in this paper.
Sect.~\ref{subsec:modelA} presents the procedure we use to derive the \stageA\ maps using the framework in Sect.~\ref{sec:model}.
Sect.~\ref{subsec:modelB} describes how we produce 
the \stageB\ maps that match the dust $TE$ correlation and $E$-$B$ asymmetry measured by \Planck, using the method 
described  in Sect.~\ref{sec:fullsky}.

\subsection{\StageA\ maps}
\label{subsec:modelA}

We explain how we produce the $(I_a,Q_a,U_a)$ maps  at a reference frequency $\nu_0$, which we choose to be $353\,$GHz, the best-suited \Planck\ channel to study dust polarization. These maps have no $TE$ correlation and no $E$-$B$ asymmetry at $\ell > 40$.

The intensity map $I_a$ is not computed from Eqs.~\ref{eq:sumIQU} but derived from observations. We use $I_a=D_{353}$, where $D_{353}$ is the dust total intensity map at $353\,$GHz 
of \cite{planck2016-XLVIII} after separation from the CIB and CMB anisotropies. 
To compute $Q_a$ and $U_a$, we need the set of angle maps $\gamma_i$ and $\psi_i$, which determine the orientation of the magnetic field in 
the $\nlay$ layers. For each layer, we draw an independent Gaussian realization for each of the three components of $\vec{\hat{B_{\rm t} }} $ in Eq.~\ref{eq:vecB}.
The angle maps are computed for the total magnetic field $\vec{B} $ including the mean magnetic field $\vec{B}_{0} $.
With the set of angles maps $\gamma_i$, using the Stokes $I$ equation in Eqs.~\ref{eq:sumIQU}, we compute the map $S_i(\nu)$ at the frequency $\nu_0$,
\begin{equation}
\label{eq:Si} 
 S_i(\nu_0)  = D_{353} / \sum^{\nlay}_{i=1} \left[1-p_0\left(\cos^2\gamma_i-\frac{2}{3}\right)\right], 
\end{equation}
where $S_i(\nu_0)$ has been assumed to be independent of the index $i$ and  $p_0$ is set to a fiducial value of 0.25. 
Next, we combine $S_i(\nu_0)$ and the angle maps $\gamma_i$ and $\psi_i$
in the Stokes $Q$ and $U$ equations in Eqs.~\ref{eq:sumIQU} to compute the ratio maps $Q_a/(p_0\times I_a)$ and $U_a/(p_0\times I_a)$ at the frequency $\nu_0$.
These ratio maps are independent of $I_a$ and depend on $p_0$ only through $ S_i(\nu_0) $.
They are computed at pixel resolution defined by the $\nside=256$ HEALPix parameter. 
After multiplication by $D_{353}$, we obtain the two maps $Q_a/p_0$ and $U_a/p_0$, which have an ill-defined beam transfer function.
The $D_{353}$ map has a resolution that varies across the sky. We 
overcome this issue by smoothing $(I_a,Q_a,U_a)$  to a resolution lower  than the lowest resolution of the $D_{353}$ map.
The model maps used in the paper have $\nside = 256$ and a symmetric Gaussian beam with a full width at half maximum of $30\arcmin$.

\subsection{\StageB\ maps}
\label{subsec:modelB}

From the harmonic coefficients of Eq.~\ref{eq:blm}, we compute the power spectra of \stageB\ maps at a given multipole $\ell$, as functions of $t,p_0,\rho,f,$ and $x$, where
\begin{equation}
\label{eq:defx}
x^2\eqdef {\cal E}\left[{\cal A}_{\ell}^{EE}/{\cal A}_{\ell}^{TT}\right]={\cal E}\left[{\cal A}_{\ell}^{BB}/{\cal A}_{\ell}^{TT}\right] \, ,\end{equation}
${\cal A}_{\ell}^{TT}$, ${\cal A}_{\ell}^{EE}$, and ${\cal A}_{\ell}^{BB}$ are the power spectra of \stageA\ maps and ${\cal E}\left[\cdot\right]$ is an averaging over multipoles between $\ell=60$ and $\ell=200$. When the slope of the $TT$ and polarization spectra are close to one another, the ratio $x$ is close to being independent of multipole $\ell$. Since this 
simplification approximately applies for dust emission (PXXX), we consider the ratio $x$ to be constant over the relevant multipole range. 

Assuming ${\cal A}_{\ell}^{XY}=0$ for $X\neq Y$, the ratios of Eq.~\ref{eq:ratios} can be expressed as follows:
\begin{equation}
\begin{dcases}
\label{eq:system}
R_{TT} &= \frac{z^2}{1+y^2} \\
R_{TE} &= \frac{z}{1+y^2}  \\
R_{BB} &= \frac{f^2y^2}{1+y^2}
\end{dcases} \, ,
\end{equation}
where $y\eqdef x/(p_0\rho)$ and $z\eqdef t/(p_0\rho)$. When the ratios $R_{XY}$ are chosen, then the system~\ref{eq:system} becomes a system of equations in $\lbrace f,y,z \rbrace$. Although the system is not linear, it can be inverted, as long as $R_{TT} > R_{TE}^2$, i.e., $\mathrm{Det}\left(\begin{array}{cc}{\cal D}_\ell^{TT} & {\cal D}_\ell^{TE} \\ {\cal D}_\ell^{TE} & {\cal D}_\ell^{EE}\end{array}\right) > 0$. When choosing values for the ratios $R_{XY}$, this condition has to be satisfied because the power spectra form a covariance, which must be positive definite. Restricting the set of solutions to positive reals, there is a unique solution, i.e., 
\begin{equation}
\begin{dcases}
\label{eq:sol}
f &=  \sqrt{R_{BB}R_{TT}/\left(R_{TT}-R_{TE}^2\right)} \\
y &= \sqrt{\left(R_{TT}-R_{TE}^2\right)/R_{TE}^2} \\
z &= R_{TT}/R_{TE}
\end{dcases} \, .
\end{equation}

From the solution of Eq.~\ref{eq:sol} and the normalization factor $N_B = (p_0f)^2$, we can compute the parameters $\rho,f,t,$ and $p_0$ as follows:
\begin{equation}
\label{eq:afpt}
f=f,\quad p_0=\sqrt{N_B}/f,\quad \rho=x/(p_0y), \quad t=zp_0\rho \, .
\end{equation}
We note that $\rho$ and the correlation coefficient between the $T$ and $E$ parts of \stageB\ maps, noted $r_{TE}=R_{TE}/R_{TT}^{0.5}$, are related as follows:
\begin{equation}
\label{rhor}
\rho = \frac{x}{p_0}\sqrt{\frac{r_{TE}^2}{1-r_{TE}^2}}
.\end{equation}

We choose the $BB$ normalization factor $N_B$ such that the power spectrum ${\cal A}_\ell^{BB}$ of \stageA\ divided by $p_0$ map is adjusted to the fit of the power spectrum 
measured over the region LR33 in PXXX (noted $C_{\ell}^{BB,data}$). Following the notation of PXXX, we have $\ell(\ell+1)C_\ell^{BB,data}=2\pi\,A^{BB,data}(\ell/80)^{\alpha_{BB}^{data}+2}$, where the values of the parameters $\alpha_{BB}^{data}$ and $A^{BB,data}$ are taken from Table~\ref{tab:ratios}. In the case where $N_B$ is $\ell$-independent, $N_B$ is the solution of the minimization of the following chi-squared
\begin{equation}
\chi^2(u) = \sum_{\ell=\ell_1}^{\ell_2} \left( {\cal A}_\ell^{BB}-\frac{1}{u} \; C_\ell^{BB} \right)^2 / \sigma_{\ell}^2\, ,
\end{equation}
with $\sigma_{\ell}^2$ the variance of $ {\cal A}_\ell^{BB}$, estimated from Monte Carlo simulations and $(\ell_1,\ell_2)=(60,200)$ as for $x$. The fit also provides the standard deviation on the normalization factor $N_B$.


\subsection{Summary of the procedure}
\label{subsubsec:proc}

The following points sketch the procedure to produce our simulations:
\begin{enumerate}
\item Draw Stokes maps $Q_a$ and $U_a$ divided by $p_0$ as described in Sect.~\ref{subsec:modelA} 
\item Mask the Galactic plane and compute the harmonic coefficients $a_{\ell m}$\item Given a mask, compute the full sky power spectra $\cal{A}_\ell$
\item Evaluate $x$ as defined in Eq.~\ref{eq:defx} and the $BB$ normalization $N_B$
\item Choose values for the ratios $R_{XY}$ of Eq.~\ref{eq:ratios}
\item Compute the corresponding solutions $\lbrace f,y,z \rbrace$ of Eqs.~\ref{eq:sol}\item Compute the parameters $\rho,f,t,$ and $p_0$ of Eqs.~\ref{eq:afpt}
\item Construct the harmonic coefficients $b_{\ell m}$ according to their definition of Eqs.~\ref{eq:blm}
\item Transform the $b_{\ell m}$'s to $(I_b,Q_b,U_b)$
\end{enumerate}

The \stageB\ maps thus constructed feature the desired two-point statistics on the desired region of the sky. The procedure can be applied on separate multipole bins, which then gives scale-dependent parameters.

\newcommand{\vecn}{\vec{n}}     
\newcommand{\vecnp}{\vec{n}^\prime}

\section{Decorrelation due to a variable spectral index}
\label{Appendix:B}

This appendix shows how to compute the decorrelation in harmonic space between two frequency maps, when spectral
differences about a mean SED may be parametrized with a spatially varying spectral index. This appendix restricts the proof to the simple case where the map that is scaled 
through frequencies and the spectral index map are correlated Gaussian white noise maps.

Let $f(\vecn)$ and $\delta\beta(\vecn)$ be two Gaussian random fields on the sphere such that
\begin{eqnarray}
\left\langle f(\vecn) \right\rangle & = & \left\langle \delta\beta(\vecn) \right\rangle = 0 \; , \\
\left\langle f(\vecn)f(\vecnp) \right\rangle & = & \delta(\vecn-\vecnp)\, \sigma_f^2 \; , \\
\left\langle \delta\beta(\vecn)\delta\beta(\vecnp) \right\rangle & = & \delta(\vecn-\vecnp)\, \sigma_\beta^2 \; , \\
\left\langle f(\vecn)\delta\beta(\vecnp) \right\rangle & = & \delta(\vecn-\vecnp)\, r\sigma_f\sigma_\beta \; .
\end{eqnarray}
From $f(\vecn)$ and $\delta\beta(\vecn)$ we construct a set of maps at frequencies $\nu_i$,
\begin{equation}
\label{eq:f_i}
f_i(\vecn) = K_i f(\vecn) \left(\frac{\nu_i}{\nu_0}\right)^{\delta\beta(\vecn)} \, ,
\end{equation}
where $\nu_0$ is a reference frequency and $K_i$ possibly contains the mean SED and unit conversion factors.

The aim is to compute the cross-spectrum $C_\ell^{\nu_i\times \nu_j}$ ($\ell\geqslant 1$) between the different $f_i(\vecn)$, i.e.,
\begin{eqnarray}
\label{eq:cl=cov}
C_\ell^{\nu_i\times \nu_j} &=& \int\mathrm{d}\vecn\,\mathrm{d}\vecnp\, \left\langle f_i(\vecn) f_j(\vecnp) \right\rangle Y_{\ell m}^\ast(\vecn) Y_{\ell m}(\vecnp) \\
 &=& \left\langle f_i(\vecn) f_j(\vecn) \right\rangle \, ,
\end{eqnarray}
where the $Y_{\ell m}(\vecn)$ represent the spherical harmonics; we assumed that two directions of the maps are uncorrelated and that the maps are statistically isotropic. 
We can rewrite the product $f_i(\vecn) f_j(\vecn) $ as follows:
\begin{equation}
\label{eq:fimix}
f_i(\vecn) f_j(\vecn) = \kappa_{ij} \left(g(\vecn) \exp\left[\frac{1}{2}\delta\gamma_{ij}(\vecn) \right] \right)^2 \; ,
\end{equation}
where $\kappa_{ij}=K_iK_j\sigma_f^2$, $g(\vecn)=f(\vecn)/\sigma_f$ and $\delta\gamma_{ij}(\vecn) = \sigma_{ij} \delta\beta(\vecn) / \sigma_\beta$ with $\sigma_{ij}=\log(\nu_i\nu_j/\nu_0^2) \, \sigma_\beta$. It can be easily verified that
\begin{equation}
\left(\begin{array}{c} g(\vecn)\\\delta\gamma_{ij}(\vecn)\end{array}\right) \sim \mathcal{N}\left( \left[\begin{array}{c}0\\0\end{array}\right], \left[\begin{array}{cc}1&r\sigma_{ij}\\r\sigma_{ij}&\sigma_{ij}^2\end{array}\right] \right)
\end{equation}
so that the expression in brackets of Eq.~\ref{eq:fimix} has a normal lognormal mixture distribution as parametrized in, for example, \cite{NLogN}. Thus, 
\begin{equation}
\left\langle f_i(\vecn) f_j(\vecn) \right\rangle = \kappa_{ij} \exp\left(\frac{\sigma_{ij}^2}{2}\right) \left(1+r^2 \sigma_{ij} ^2\right)
\end{equation}
and
\begin{equation}
\begin{split}
\frac{C_\ell^{\nu_i\times\nu_j}}{\sqrt{C_\ell^{\nu_i\times\nu_i}C_\ell^{\nu_j\times\nu_j}}} = & \exp\left\lbrace-\frac12\sigma_\beta^2\left[\log\left(\frac{\nu_i}{\nu_j}\right)\right]^2\right\rbrace \\
 & \times \frac{\left(1+r^2 \sigma_{ij} ^2\right)}{\sqrt{\left(1+r^2 \sigma_{ii} ^2\right)\left(1+r^2 \sigma_{jj} ^2\right)}}.
\end{split}
\end{equation}
We note that the correlation does not depend on the mean SED.

\end{document}